\documentclass{aa}
\usepackage{graphicx}
\usepackage{comment}
\usepackage{multirow}
\usepackage{upgreek}
\usepackage{pifont}
\usepackage{pdfpages, caption}
\usepackage[bookmarks,bookmarksnumbered,%
            urlcolor=blue,citecolor=blue,linkcolor=blue,%
             colorlinks=true,hyperfigures=true,hyperindex=true]{hyperref}
  
\usepackage{xcolor}
 \usepackage{color}

\definecolor{orange}{rgb}{1,0.5,0}

\definecolor{cadmiumgreen}{rgb}{0.0, 0.42, 0.24}

\newcommand{\msun}{M_{\odot}}
\newcommand{\reros}{R_\mathrm{eros}}
\newcommand{\beros}{B_\mathrm{eros}}
\newcommand{\rerosccd}{R_\mathrm{eros1}^\mathrm{CCD}}
\newcommand{\berosccd}{B_\mathrm{eros1}^\mathrm{CCD}}
\newcommand{\Bj}{V_{\rm J}}
\newcommand{\Vj}{V_{\rm J}}
\newcommand{\Rc}{R_{\rm C}}
\newcommand{\Ic}{I_{\rm C}}
\newcommand{\erosII}{EROS-2}
\newcommand{\erosI}{EROS-1}
\newcommand{\erosIccd}{EROS-1-CCD}
\newcommand{\erosIplates}{EROS-1-Schmidt-Plates}
\newcommand{\apxbf}[1]{#1}

\usepackage{txfonts}
\begin{document} 

   \title{EROS light curve data base} 

\author{
C. Afonso\inst{1},
J.-N. Albert\inst{2},
R. Ansari\inst{3},
E. Aubourg\inst{1,4},
J.-P. Beaulieu\inst{5},
T.~Blaineau\inst{2},
C. Coutures\inst{1},
F. Derue\inst{7},
J.-F. Glicenstein\inst{1},
B. Goldman\inst{6},
C. Hamadache\inst{2},
T. Lasserre\inst{1},
L. Le Guillou\inst{7},
E. Lesquoy\inst{1},
C. Magneville\inst{1},
B. Mansoux\inst{2},
M.~Moniez\inst{2},
N. Palanque-Delabrouille\inst{1},
O. Perdereau\inst{2},
J. Rich\inst{1},
M. Spiro\inst{1},
P. Tisserand\inst{5}
}
\institute{
IRFU, CEA, Universite de Paris-Saclay, CEA, Irfu, 91191 Gif-sur-Yvette, France
\and
Université Paris-Saclay, CNRS/IN2P3, IJCLab, 91405 Orsay, France
\and
Université Paris-Saclay, Université Paris Cité, CEA, CNRS, AIM, 91191, Gif-sur-Yvette, France
\and
Universit\'e de Paris, CNRS, Astroparticule et Cosmologie,  F-75013 Paris, France
\and
Sorbonne Universit\'es, UPMC Univ Paris 6 et CNRS, UMR 7095, Institut d'Astrophysique de Paris, IAP, F-75014 Paris, France
\and
Universit\'e de Strasbourg, CNRS, Observatoire Astronomique, CNRS, UMR 7550,F-67000, Strasbourg, France
\and
Sorbonne Universit\'e, CNRS/IN2P3,
Laboratoire de Physique Nucl\'eaire et de Hautes \'Energies (LPNHE),
75005 Paris, France
}

\offprints{J. Rich,\hspace{2mm} \email{ james.rich@cea.fr}}

   \date{Received ...2025; accepted ...2025\\}
\authorrunning{EROS coll.}
\titlerunning{The EROS light-curve database}

  \abstract
 { 
 The EROS project ({\sl Exp\' erience de Recherche d'Objets Sombres })
 carried out 
 photometric surveys of dense stellar fields towards the Magellanic Clouds (LMC and SMC), the Galactic Bulge  and Galactic spiral arms, over the period 1990-2003. 
 The main goal of the experiment was to search for the Galactic Dark Matter in the form of massive compact objects (machos), through the gravitational microlensing effect. 
 }
{
The historical record of the flux variations of the monitored stars by EROS-2 will be a unique asset for time domain astronomy and to complement current and future searches of transient sources.
}  
{
We describe the set of light curves obtained by EROS-2 program over the years 1996 to 2003, monitoring more than 86 million stars, which is publicly released through the Centre de Donn\'ees de Strasbourg (CDS).
  The numbers of light curves in this data set are 
28.7 and 4.0 million in LMC, respectively SMC, 42.9 million in the Galactic Bulge and 
10.4 million towards the Galactic spiral arms,
with several hundred measurements for each object. Data from EROS-1 is also being released. 
}
{
 The object catalog and light curves and images  are accessible through the CDS portal.
 This will be useful for checking the  past behavior   of newly discovered variable objects.
}


   \maketitle
   \nolinenumbers
%

\section{Introduction}
\label{section:1intro}
The EROS\footnote{EROS \href{http://eros.in2p3.fr}{http://eros.in2p3.fr}} (Expérience de Recherche d'Objets Sombres) project started in 1990 \citep{1993Msngr..72...20A} with the aim of searching for massive compact objects, co-called machos, such as brown dwarfs or black holes that could make  up a significant fraction of the Milky Way dark matter halo. 
The project was inspired by the suggestion of B. Paczynski in 1986, \citep{1986ApJ...304....1P}, that machos could reveal their presence 
through the gravitational microlensing effect, acting on background stars. 
The name microlensing was coined by B. Paczynski, to emphasize on the tiny angular separation of 
a fraction of milli-arcsec between multiple images. 
However, it can be observed as a transient 
magnification of the background star due to the relative observer-lens-star motion.  The duration of the transient star brightness increase
depends on the geometric configuration and scales with
the square root of the lens mass. It is typically 70 days for $1\msun$ lenses in the halo of the Milky Way, and sources in the Magellanic clouds. 

The two phases of EROS, EROS-1 and EROS-2, observed the Magellanic clouds from the European Southern Observatory (ESO) at La~Silla, Chile, over the period 1991 through 2003.
Two other microlensing projects were initiated nearly simultaneously with EROS: the MACHO project observing from Mount Stromlo in Australia, and the OGLE project observing from the Las Campanas observatory in Chile. The first candidate events were announced by the three projects in 1993 \citep{Alcock1993,1993Natur.365..623A,Ogle1993}.
These three microlensing projects were later joined by the MOA project \citep{MOA_CG2003}, who focused primarily on the search for planetary systems through their microlensing effect, leading to a first planet candidate reported in \citep{2004ApJ...606L.155B}.  

After 5.7 years of observations,
the MACHO collaboration reported a significant number of events toward the central regions of the LMC \citep{2000ApJ...542..281A}, 
which was interpreted as being due $\sim 0.5 M_\odot$ machos, making up $\approx 20\%$ of the standard halo total mass. 
Subsequent results from EROS \citep{2007A&A...469..387T} and OGLE \citep{ogle_2011} cast doubt on the interpretation of the MACHO events as due to halo objects by limiting the macho mass fraction of the halo to less than $\approx10\%$ over the macho mass range $10^{-6}$ to $10M_\odot$. 
A new analysis using a combined data set of EROS and MACHO light curves extended these limits up to $1000M_\odot$ \citep{2022A&A...664A.106B}.  
More recently, the OGLE results have further
lowered these limits to the 1\% level \citep{Ogle_2024Natur.632..749M, Ogle_2024ApJS..273....4M, 2024ApJ...976L..19M,2025ApJS..280...49M}.


The first phase of EROS, EROS-1, observed the Magellanic clouds between 1991 and 1995,
using two complementary instrumental setups.
The first used
ESO Schmidt photographic plates to search for events of duration more than a few days. The final results of this program were presented in \citet{1996A&A...314...94A}.
The second setup used the 40 cm aperture, T40 telescope equipped with the EROS-1 CCD camera to search for events with durations as short as one hour.
\citep{1994ExA.....4..265A,1994ExA.....4..279A}.
The final results of this program were presented in  \citet{1998A&A...329..522R}

The second phase, EROS-2, observed between 1996 and 2003. It used the 1 meter MARLY telescope, equipped with  two CCD cameras, each with 64 million pixels, to perform a systematic survey from 1996 to 2003 
of seven targets described in Table \ref{tab:listoftargets}:
the two Magellanic clouds (LMC, SMC), Galactic Centre (CG) and four regions in the direction of Galactic spiral arms ($\beta$Sct, $\gamma$Sct, $\gamma$Nor, $\theta$Mus). 
The final microlensing results using bright stars in the LMC and SMC were presented in \citet{2007A&A...469..387T}.
The results for the Galactic center using clump giants were presented in \citet{Hamadache2006}.  The optical depth toward the spiral arms was given in \citet{Rahal_2009}.

In this paper, we present the EROS-2 catalogs and light curves of 85 million stars,
including  variable stars and faint stars not used for measurements of the optical depth.
The catalogs are being made public through the Centre de Donn\'ees astronomiques de Strasbourg (CDS)\footnote{CDS, Strasbourg astronomical Data Center: \href{https://cds.unistra.fr}{https://cds.unistra.fr}}, 
so that they can be consulted by ongoing or future major surveys programs, such as Legacy Survey of Space and Time 
(LSST)\footnote{LSST is the survey program that will begin at the end of 2025 by the Vera C. Rubin Observatory : \href{https://rubinobservatory.org}{https://rubinobservatory.org}}, 
to check the past history of transient objects that will be discovered.  
Brokers such as Fink\footnote{Fink: The transient alert broker system associated with LSST - \href{https://fink-broker.org/about/}{https://fink-broker.org/about/} } \citep{2021MNRAS.501.3272M} will also be able 
to use the EROS light-curve database  to confirm or deny alerts, particularly in the search for microlensing events.

The paper is organized as follows.
The EROS-2 instrument and observing program are  presented in section \ref{section:2eros2} and
the image processing pipeline in section \ref{section:3dataprocessing}.
The data quality is discussed in section 
\ref{section:quality}. The characteristics of the public data set is given in section \ref{section:dataset} with the detailed data structure,
illustrated in Fig. \ref{fig:file_examples}, given in appendix \ref{section:fielddesc}.
The \erosII \, images being released are described in appendix \ref{sec-eros2-images}.
The \erosIccd~data being made public is briefly described in appendix \ref{sec-eros1-data}
and that of the EROS-1 Schmidt-plate program in appendix \ref{sec-eros1-schmidtplate-data}. 

\begin{table*}
\begin{center}
\caption{Main characteristics of the EROS~microlensing targets. \label{tab:listoftargets}}
\scalebox{0.85}{\begin{tabular}{|c|c|c|c|c|c|c|c|c|c|c|}
\hline
target & r.a. range & dec. range & $N_{fields}$ & $N_{exposures}$ & $N_{stars}$ & $N_{stars}$ & $\beros$ & $\reros$  & $\beros$ & $\reros$ \\ 
      &  (deg) & (deg)  &  &  & $(10^6)$ &  $\sigma_B<0.1$ & $\sigma_B \approx 0.1$ & $\sigma_R\approx0.1$ & 90\% & 90\% \\
\hline
\erosII & & & & & & & & & & \\
LMC (lm) &  69.79 : 98.54  &  -76.06 : -63.46  &  88  &  44706  &  28.72  &  5.79  &  19.4  &  18.6  &  21.8  &  21.4 \\
SMC (sm) &  6.39 : 19.03  &  -74.8 : -71.52  &  10  &  8614  &  4.03  &  0.87  &  19.6  &  18.7  &  22.1  &  21.7 \\ 
CG (cg) &   259.3 : 279.51  &  -37.19 : -17.02  &  81  &  35644  &  42.09  &  16.22  &  18.7  &  17.9  &  20.3  &  19.2 \\  
$\beta$Sct (bs) & 280.51 : 283.49  &  -8.38 : -5.23  &  6  &  2172  &  2.25  &  0.67  &  18.9  &  18.1  &  21.4  &  20.0 \\
$\gamma$Sct (gs) & 276.68 : 279.0  &  -15.56 : -12.12  &  5  &  1819  &  1.8  &  0.49  &  19.1  &  18.3  &  21.6  &  20.2 \\
$\gamma$Nor (gn) &  241.79 : 248.47  &  -55.88 : -51.08  &  12  &  7207  &  4.26  &  1.03  &  18.8  &  18.1  &  21.6  &  20.3 \\
$\theta$Mus (tm) &  197.2: 204.49  &  -65.73 : -62.38  &  6  &  3078  &  2.13  &  0.58  &  18.6  &  17.8  &  21.3  &  19.9 \\
\hline
\erosI & & & & & & & & & &\\
Schmidt & 71.90: 87.99 & -71.18: -65.79 & 1 & 249 & 7.82 & 0.34 & 18.0 & - & 21.7 & 20.6 \\
CCD  LMC& 78.84: 82.12 & -69.88: -69.25 & 1 & 10984 & 0.16  & 0.06 &  18.5 & 18.6 & 20.2 & 19.7 \\
CCD  SMC & 10.73: 14.75 & -73.53: -72.91 & 1 & 5626  & 0.15 & 0.04 & 18.8 & 18.5 & 20.8 & 20.6 \\
\hline
\end{tabular}
}
\tablefoot{
The EROS~microlensing targets (column 1) with the ranges in right ascension and
declination, the number of fields, the number of exposures, the total number of catalogued stars,
and the number of stars with photometric precision in $\beros$ less than 0.1mag.
Columns eight and nine give the approximate EROS magnitudes corresponding to a precision
of 0.1mag and columns 10 and 11 give the approximate magnitude limits containing  90\% of the stars.   The precise values of \erosII~magnitudes are field-dependent because of differences in exposure time and stellar density.
The magnitudes for the \erosIplates~magnitudes are for objects outside the LMC bar.
}
\end{center}
\end{table*}

\begin{figure*}[htbp]
  \centering
  \includegraphics[width=0.5\textwidth]{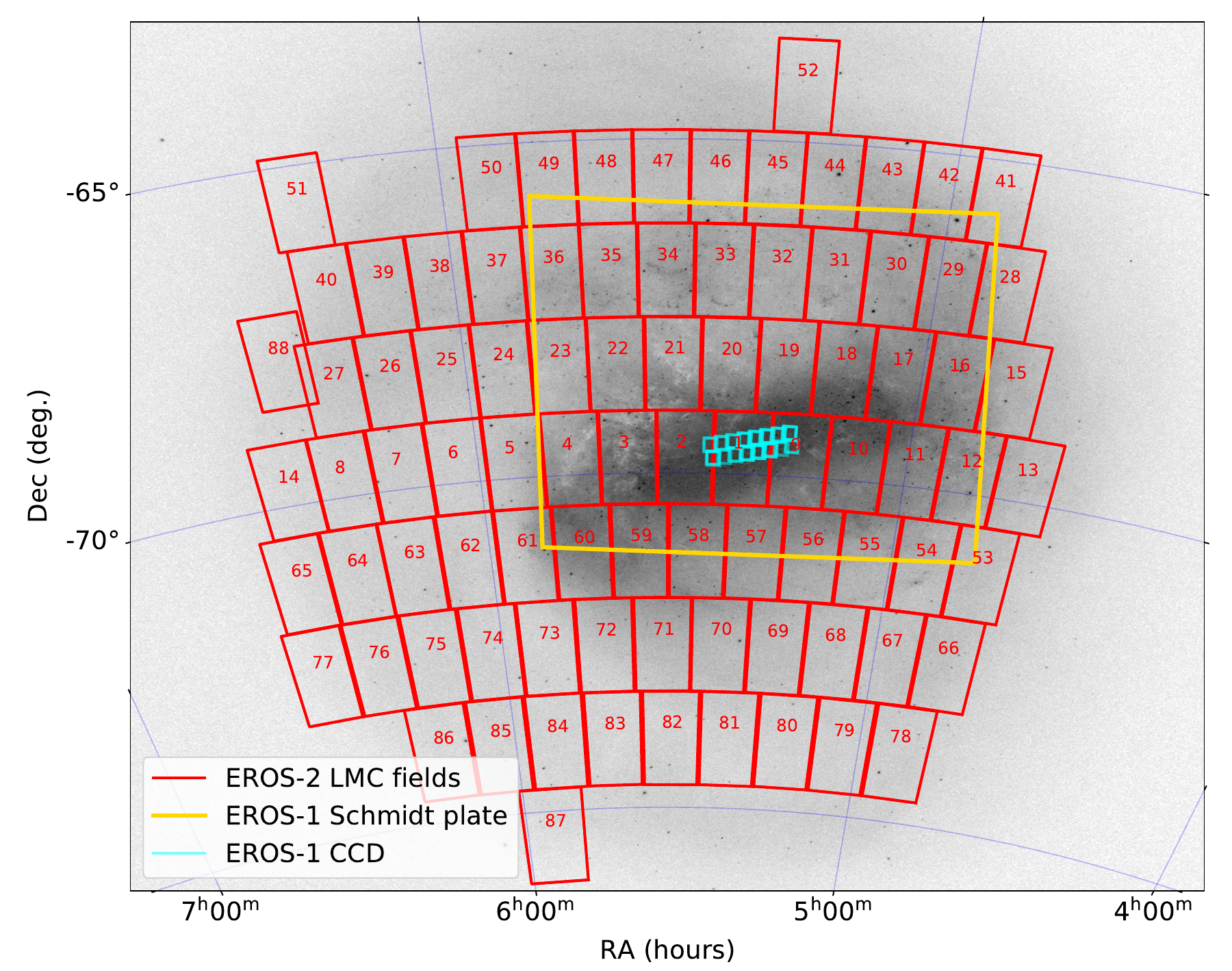}
  \hspace{5mm} 
\includegraphics[width=0.35\textwidth]{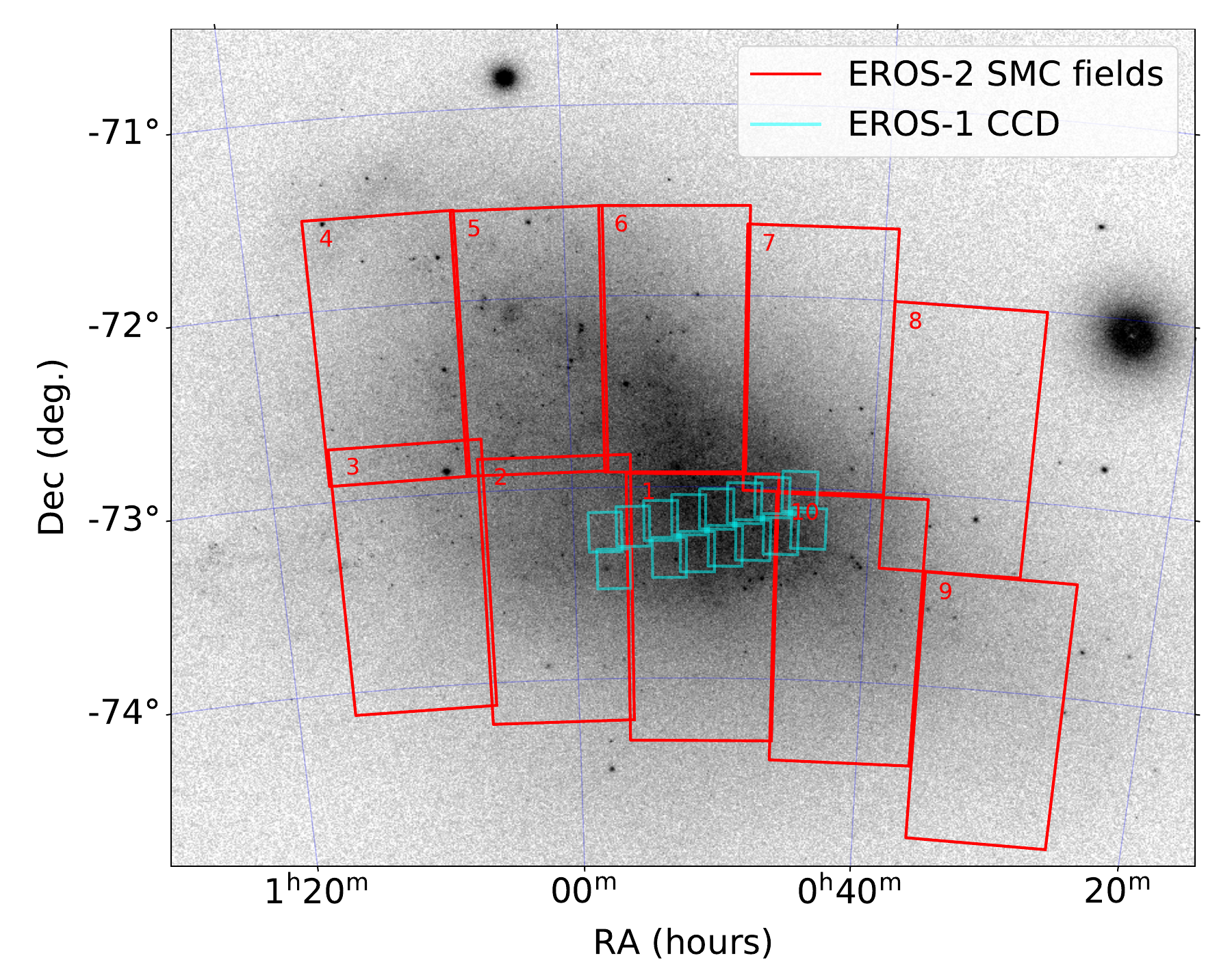} 
\caption{The EROS LMC (left) and SMC (right) fields superimposed on images of the full 
 {\it Gaia-dr3} catalog numbers counts  \citep{gaia_2023A&A...674A...1G}, 
 extracted using {\tt hipstofits}.  
The \erosII~fields are shown as solid red lines, while the 
\erosI fields are shown in cyan for the survey with the T40/CCD camera and in yellow for the LMC Schmidt plates. 
}
  \label{fig:erosfieldsmap}
\end{figure*}

\begin{figure*}[htbp]
  \centering

 \includegraphics[width=\textwidth]{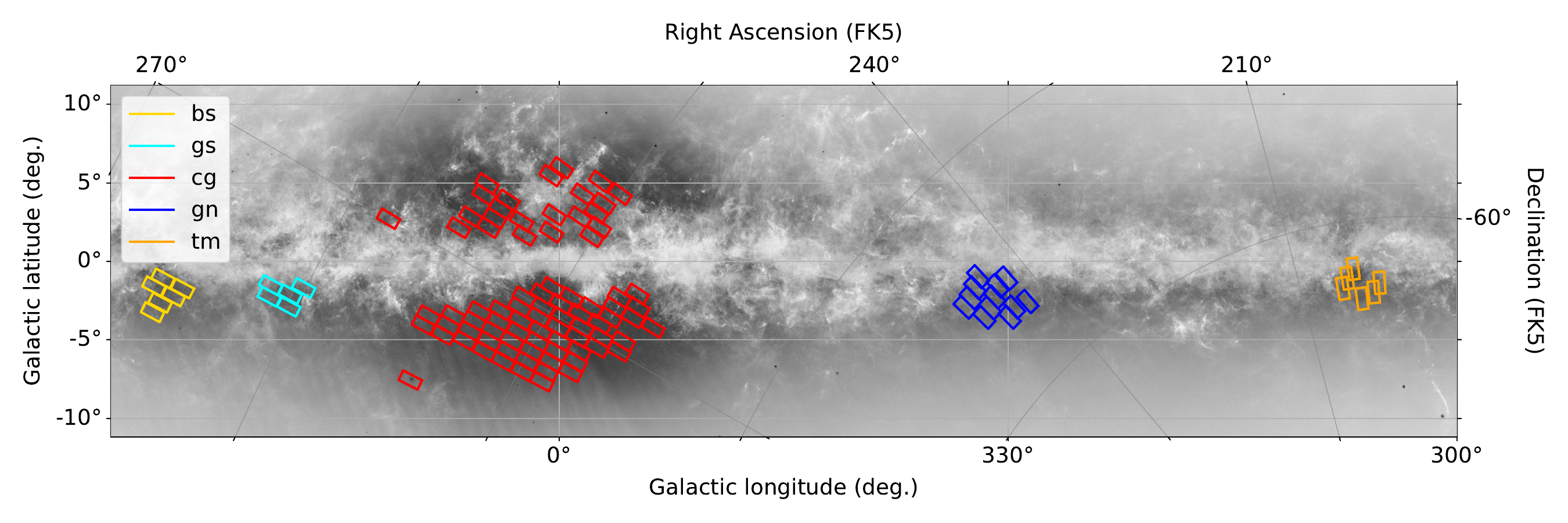}\\

\caption{Map of the Galactic-plane fields regularly observed by EROS-2.
The colored contours correspond to the EROS-2 fields, color-coded for the 5 regions. The background image shows the density of the full {\it Gaia-dr3} sources \citep{gaia_2023A&A...674A...1G}, with a square root scale,  as extracted  from the CDS using  
{\tt hips2fits}. 
Some  artifacts in the form of stripes can be seen in the Galactic bulge  area, reflecting Gaia's scanning strategy impact on its source detection threshold.
}
  \label{fig:erosgpfieldsmap}
\end{figure*}

\section{EROS-2 program}
\label{section:2eros2}

\subsection{The instrument}

A brief overview of the EROS-2 telescope and camera is given here. 
A detailed presentation  of the instrument, 
telescope operations and data reduction can be found in   
in \citet{bau97}, \citet{bau_1997NIMPA.387..286B}, and \citet{1997PhDT.........2B}
and references therein.

\begin{figure}[htbp]
  \centering
  \includegraphics[width=0.35\textwidth]{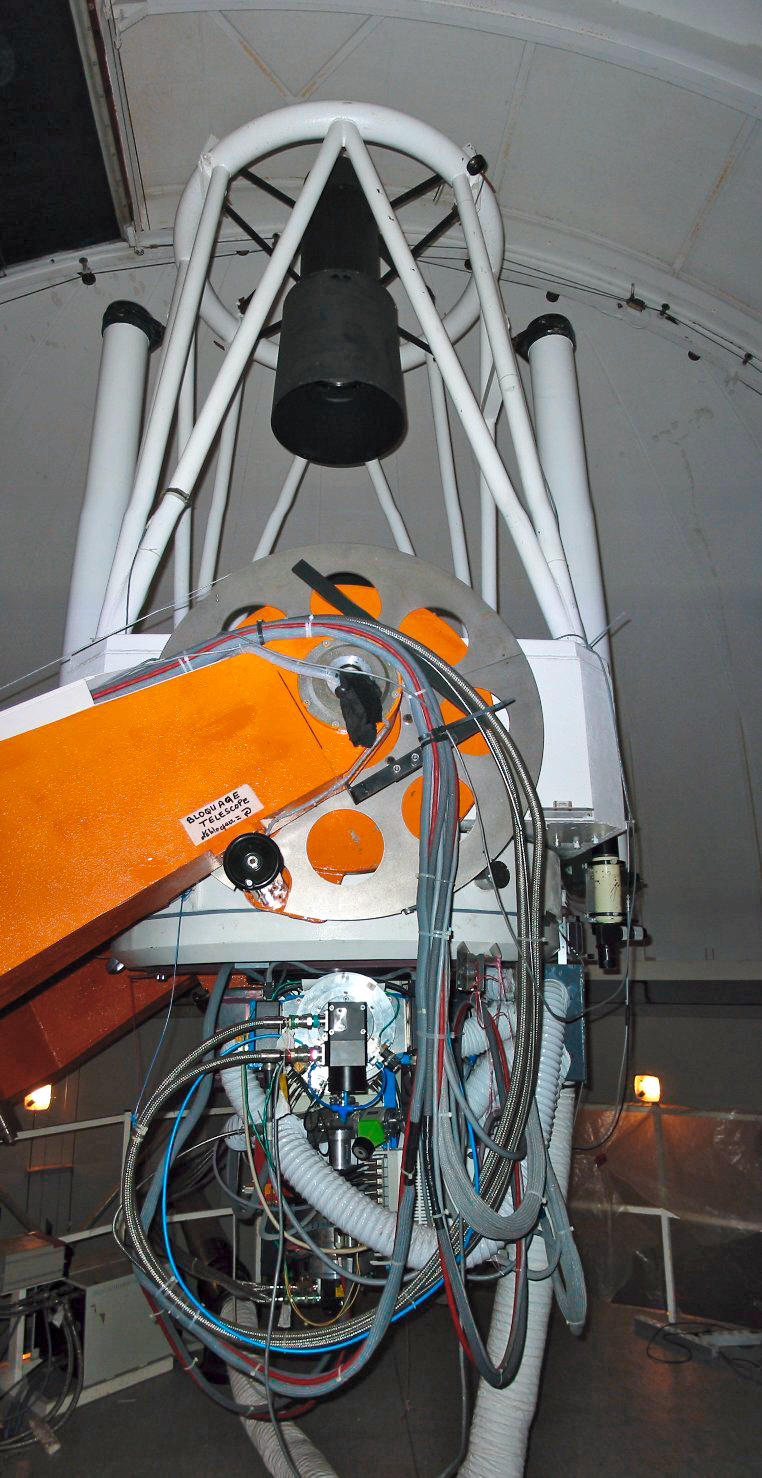}
  \caption{The MARLY telescope, equipped with two EROS CCD cameras.}
  \label{fig:marlycamera}
\end{figure}

The EROS-2 project used a $96\;\mathrm{cm}$ Ritchey-Chr\'etien telescope named MARLY.
The telescope was originally designed and built in 1970 by Observatoire de Marseille and Observatoire de Lyon. It was refurbished and  installed in the GPO dome (Grand Prism Objectif) in the ESO La Silla observatory in Chile in 1995\footnote{MARLY telescope at ESO : \href{https://www.eso.org/public/teles-instr/lasilla/marly/}{https://www.eso.org/public/teles-instr/lasilla/marly/}}. Figure~\ref{fig:marlycamera} shows the telescope instrumented with the two EROS cameras.
The telescope was equipped with a focal reducer and field corrector, decreasing the focal ratio from $f/8$ to $f/5$ to increase the field of view, leading to a focal length $f=~5.14 \mathrm{m}$. 
A dichroic beam-splitter was also installed with a separation around $H_\alpha$ line at 656 nm, dividing the incoming light between a blue and a red arm, each equipped with a wide $\sim 1 \mathrm{deg^2}$ field of view camera.

\begin{figure*}[htbp]
  \centering
  \includegraphics[width=0.35\textwidth]{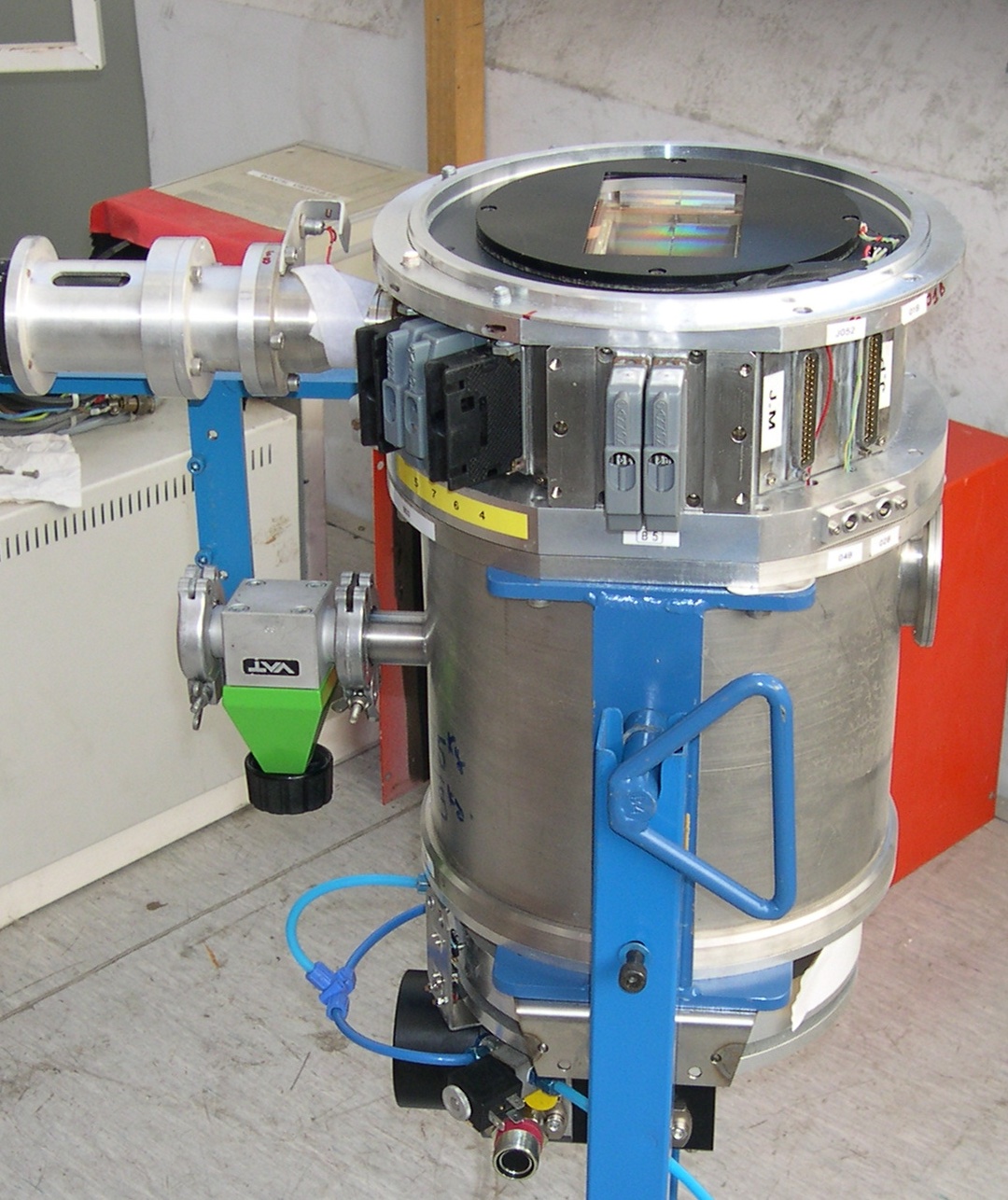}
  \hspace{5mm}
  \includegraphics[width=0.55\textwidth]{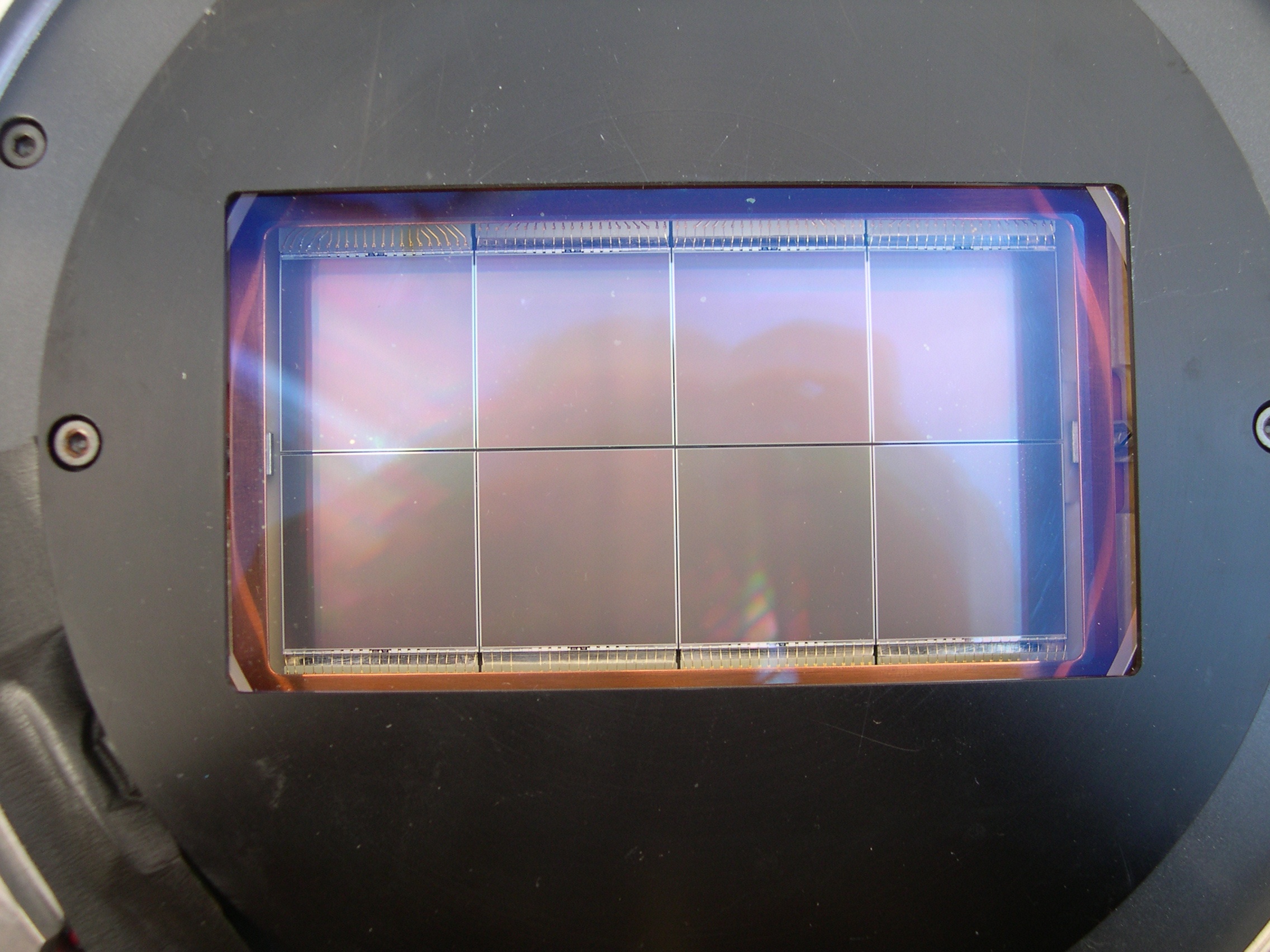}
  \caption{Left: The EROS-2 blue camera cryostat. Right: a view of the mosaic, showing $2 \times 4$ array of $2048 \times 2048$ CCD sensors.}
  \label{fig:cryomosaic}
\end{figure*}

Each camera had an array of $2 \times 4 = 8$ CCD's 
with the 4 adjacent CCD's aligned with the declination axis. The CCD sensors had 2048~x 2048 pixels of $15 \times 15\, \mathrm{\mu m^2}$ size, corresponding to 
$0.602 \times 0.602$ arcsec$^2$.
The mosaic of 8 CCD's in each camera, with a total of 32 millions pixels ($8K \times 4K$) covered a $120 \times 60 \mathrm{mm^2}$ area on the focal plane, corresponding to  $0.7\degr \times 1.4\degr$ on sky. 
The three sided buttable $\mathrm{2K \times 2K}$ Loral\footnote{
The Loral corporation was a defense contractor founded in 1948 and acquired by Lockheed-Martin in 1996. Check \href{https://en.wikipedia.org/wiki/Loral\_Corporation}{https://en.wikipedia.org/wiki/Loral\_Corporation} for the Loral corporation history and \href{https://www.inknowvation.com/sbir/story/our-history-fairchild-imaging}{https://www.inknowvation.com/sbir/story/our-history-fairchild-imaging} for the CCD manufacturing activities. An overview of the history of the development and use of large CCD array sensors in astronomy at the time of EROS-2 camera design can be found in \citep{1992ASPC...23....1J}. }
$2K3eb$ CCD's were packaged at the Steward Observatory (University of Arizona),
then tested and assembled into the two mosaics at the CEA DAPNIA Institute (IRFU today) in Saclay, France.
The wavelength pass-band including the CCD quantum efficiency was $400-735\;\mathrm{nm}$ for the blue camera and $550-950\;\mathrm{nm}$ for the red camera.
The CCD mosaics were cooled down to $180\;\mathrm{K}$ in two vacuum vessels using a $\mathrm{He}$ Edwards cryocooler. Photographs showing one of the cryostats and a closeup view of the its mosaic is shown on Figure \ref{fig:cryomosaic}.

The CCD clocking and readout was handled by a custom designed electronic system, achieving a per-pixel RMS noise of 6-7 electrons. 
The VME (Versa-Module Europe) based acquisition system was  equipped with MMB (Memory Management Boards) and controlled the parallel readout of the two cameras, with a readout time of 50 seconds.
The image data was then transferred to unix (DEC-OSF) acquisition computers through Ethernet network, then saved as FITS files. Each exposure produced  $2 \times 64 \mathrm{MB}$ or 128 megabytes of data, representing a total of 6-8 gigabytes image data per night.  

An automatic telescope pointing and guiding system was designed by CNRS/IN2P3 LAL (Laboratoire de l'Acc\'elerateur Lin\'eaire)
in Orsay, now IJCLab. The system featured a custom designed control and supervisor software which ran on a standard PC, with Linux operating system; its operation and performances could be monitored remotely, from France. The telescope angular positions on the two rotation axis were obtained through commercially available on-axis absolute encoders and associated read-out electronics, and each axis was also equipped with precision electrical motors, which were driven by a motion control board with PID filters, producing the necessary torque to move the telescope. The dome positioning was also automatic and piloted by the same software.
Finally, this system used a 1600x800 pixels CCD camera, observing a small sky area picked on the side of the field of view of the main cameras by a specially designed optics. The system was able to move the telescope to the specified sky coordinates, find and lock the guide star while the CCD mosaics was being readout, thus introducing minimal dead time during the intensive observations.
The choice of using a CCD camera for the EROS-2 guiding system was motivated in \citet{mansoux1997} where the performance of the system was studied for CCD camera and compared to that of an intensified camera.  

A first processing of the exposures, for quality checks was carried out at the dome using a cluster of three DEC\footnote{The Digital Equipment Corporation was a US computer manufacturer company founded in 1957, active till the end of 1990's, before being taken over by Compaq. Check \\ \href{https://en.wikipedia.org/wiki/Digital\_Equipment\_Corporation}{https://en.wikipedia.org/wiki/Digital\_Equipment\_Corporation} or \\ \href{https://www.computerhistory.org/brochures/d-f/digital-equipment-corporation-dec/}{https://www.computerhistory.org/brochures/d-f/digital-equipment-corporation-dec/} \\ for more information about the company history and products.} Alpha $600$ computers.
The data were stored on DLT tapes and sent to the CC-IN2P3 (Centre de Calcul de l'IN2P3) in France.

\begin{figure}[htbp]
  \centering
\includegraphics[width=0.45\textwidth]{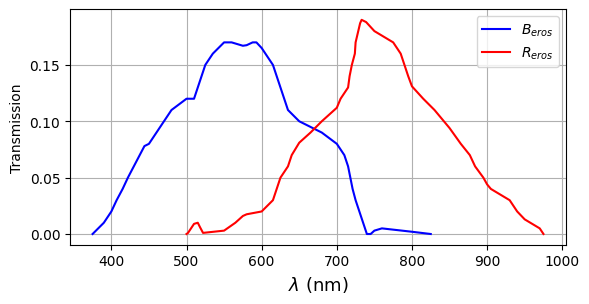}
  \caption
  {The calculated transmission of the \erosII~optics \citep{1997PhDT.........2B}.}
  \label{fig:transmission}
\end{figure}

Images were taken simultaneously in two wide passbands, $\reros$ centered close to the $\Ic$ standard band,
and $\beros$ intermediate between the standard $\Vj$ and $\Rc$ bands. 
The calculated passbands of the two arms are shown in Fig. \ref{fig:transmission}.
After the end of the EROS survey,
almost all of our  Magellanic fields could be calibrated using stars from the  catalogs of the
Magellanic Clouds Photometric Survey  \citep{zaritsky}.
For $4.5\,\deg^2$, the calibration was checked with the   OGLE-II  catalog \citep{ogleIIphotometry}.  
To a precision of $\sim0.1\,{\rm mag}$, the EROS magnitudes satisfy
\begin{equation}
\reros=\Ic \hspace*{5mm} \beros\;=\;\Vj - 0.4(\Vj-\Ic) \;.
\end{equation}

\subsection{The survey}

Between July 1996 and February 2003, EROS-2 monitored a number of dense stellar fields within the LMC, the SMC, the Galactic Centre and 4 regions also located in the Galactic plane, but far from the centre. In particular, regions of the Milky Way have been selected as having undergone the least dust extinction.

\begin{figure}[htbp]
  \centering
  \includegraphics[width=0.48\textwidth]
  {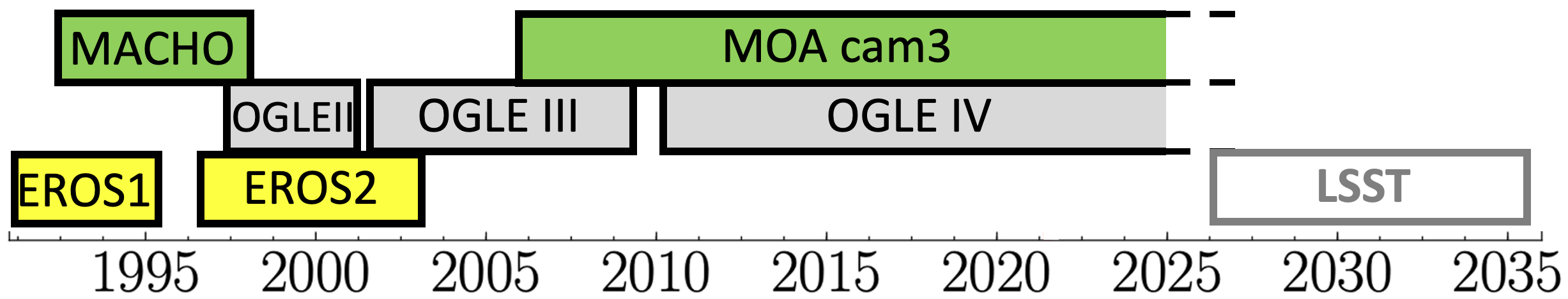}
\caption{Timeline of observation epochs for the various surveys dedicated to microlensing
}
  \label{fig:microlepochs}
\end{figure}

Table \ref{tab:listoftargets} describes the data acquired for these different programs, and Figures \ref{fig:erosfieldsmap} and \ref{fig:erosgpfieldsmap} shows the maps of the fields regularly imaged by EROS-2. Fields used to search for supernovae that have not been monitored over long periods are not shown.
Figure \ref{fig:microlepochs} shows the observation epochs covered by EROS-1 and EROS-2, and the other main microlensing surveys.

The main scientific objective of the studies towards LMC and SMC was the search for massive compact objects in the Galactic halo through microlensing. These programs found only a very small number of microlensing events, in numbers compatible with the expectations of self-lensing (due to lenses belonging to the LMC or SMC) plus lenses in the Galactic disc. The results of this program have been published and interpreted in references \citep{2007A&A...469..387T,2022A&A...664A.106B}, and concluded that there is a negligible contribution to the Galactic halo from massive compact objects with masses between $10^{-7}$ and $1000$ solar masses.

Searches toward the Galactic Centre led to the discovery of hundreds of confirmed microlensing events 
and to the publication of the optical depth map which traces the column-density of matter \citep{Hamadache2006}.

Long a specificity of EROS-2, the search in the  Galactic Spiral Arms (GSA), far from the Galactic centre, has also made it possible to constrain the contribution of the halo and, above all, to test a thick disc model. The article \citep{Rahal_2009} provided the optical depths in the 4 regions studied based on the detection of 22 microlensing event candidates.
A particular difficulty in interpreting these optical depths is that the sources are not at a single distance (as in LMC), but are distributed along the line of sight. In another interpretation article \citep{BS2017}, we have integrated the information provided by the magnitude-color diagrams of the fields studied in order to compare all the observables with the Galactic models.

Systematic searches for variable stars, both periodic and non-periodic, were also carried out in each monitored field. 
The results for searches in the Magellanic Clouds were presented in \citet{tisserand_rcb_mc_2009A&A...501..985T,Epoch_2014},
in the Galactic Bulge in \citet{tisserand_rcb_cg_2008A&A...481..673T},
and in the spiral arms in \citet{Derue_2002}.

The microlensing fields were not observable in good conditions all  the time. 
Two other major programs were therefore performed.
The first searched for high proper-motion stars and led to limits on the contribution of cool
white dwarfs to the Galactic Halo \citep{Goldman_2002}. 

The second program, comprising
about 10\% of the  observing time,  was a  semi-automated supernovae search. 
Fields in low galactic interstellar absorption areas were observed twice at 15 to 20 days intervals. An automatic image subtraction and candidate selection was then performed. After a visual scan during the day following the second observation,
a final set of potential SN candidates were subjected to complementary observations and/or spectroscopic observations for confirmation. 
In 1997-98, in total  about 35 SN were discovered. Spectroscopic observations on the ESO 3.6m and ARC 3.5 m telescopes showed than 7 of these SNe were of
type Ia, 1 of type Ic and 2 of type II. 
A measurement of the SNIa  explosion rate at $z \sim 0.1$ was extracted from these campaigns \citep{2004A&A...423..881B}.  In the first months of 1999, EROS-2  participated in an international SNIa search campaign led by the Supernova Cosmology Project (SCP). Nine out of the 19 SNIa sample gathered by this effort were discovered by EROS-2. This nearby SNIa sample was used to improve cosmological parameter determination \citep{regnault2000,2008ApJ...686..749K}.


\section{Data processing pipeline}
\label{section:3dataprocessing}

The processing of the astronomical images taken by the EROS-MARLY telescope and camera represented a major challenge in the 90's, given the generated data volume and the digital storage and computing power available at that time. Data obtained each night was copied on tape cartridges and shipped from the ESO La Silla observatory to the CNRS/IN2P3 computing centre\footnote{CC-IN2P3 : \href{https://cc.in2p3.fr/en/}{https://cc.in2p3.fr/en/} } in Lyon, where the data processing and analysis was carried out. The data access and light curve production was managed by a set of tools, ErosDb, relying on a relational database. Originally developed in Tcl, ErosDb\footnote{ErosDb : EROS data and processing management software tools - \href{https://erosdb.in2p3.fr}{https://erosdb.in2p3.fr} } has been subsequently rewritten in Java.  The image processing and photometric reconstruction was handled by the PEIDA\footnote{PEIDA : {\bf P}hotom\'etrie et {\bf E}tude d'{\bf I}mages {\bf D}estin\'e \`a l'{\bf A}strophyique - \href{http://eros.in2p3.fr/peida}{http://eros.in2p3.fr/peida} } object oriented C/C++ software, designed and developed for the  EROS project  \citep{1996VA.....40..519A}. The EROS images were stored as standard FITS format files, with one file for each of the CCD's and the light curves were stored in structured compact binary files or the suivi {\tt .sv} files, with efficient data access.    

\subsection{PEIDA}
\begin{figure}[htbp]
\begin{center}
    \includegraphics[width=0.44\textwidth]{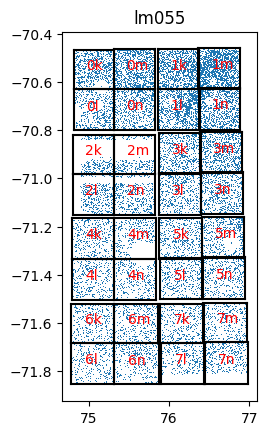}
    \caption[]{
    Map of the lm055 field showing positions of stars with $\beros<18.$ and the boundaries
    of the CCD-quarters used in star catalogs.
    The empty circles correspond to masked regions centered on bright foreground stars causing CCD saturation.
    The empty rectangles on CCD 2 and 7 are dead regions of the red CCD.
    }
    \label{fig:ccdquarterslm055}
\end{center}
\end{figure}

\begin{figure}[htbp]
\begin{center}
    \includegraphics[width=0.42\textwidth]{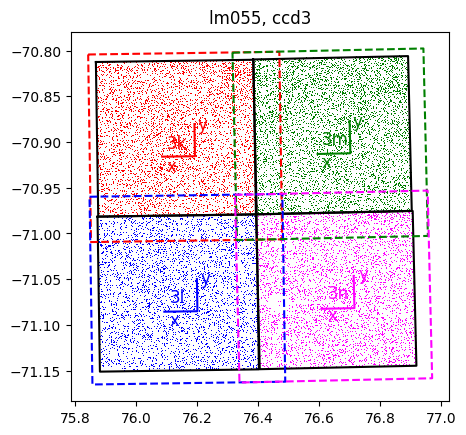}
    \caption[]{
    Zoom of CCD 3 showing the boundaries of the reference images and the orientation of the $x$ and $y$ axes.
    }
    \label{fig:ccdquarterslm055_3}
\end{center}
\end{figure}

The standard processing method used in EROS for the production of light curves relied on the
{\it Forced Photometry} paradigm. This assumes that a common set of celestial sources, with fixed positions on sky, but possibly with changing luminosity, is contributing to each image, in addition to a sky background, nearly uniform over the focal plane. The sky background is due to the light from the moon, the atmosphere, the instrument, and unresolved sources. 
This master source list, called the {\it Reference Catalog} was obtained from the detection of non extended, point like sources  on a deep, high quality image {\it (Reference or Template Image)} for each surveyed field. The light curve for each object was then reconstructed by determining the source luminosity or flux on each image, using PSF (point spread function) photometry. A light curve gathers the photometric measurements, flux or magnitude, with the associated date and time of the corresponding image or exposure, as well as ancillary information such as the photometric uncertainty or the sky background level. The main steps of the processing are briefly described below. Note that images were processed not on a per CCD basis, but rather of a quarter of CCD. 
The image processing pipeline used quarter of CCD image patches, covering $\sim 10 \times 10 \, \mathrm{arcmin}^2$ on the sky, with some overlap between quarters, for most processing steps. 
This helped mitigating the effect of parameters held constant over the processing patch such as the PSF parameters, 
while ensuring acceptable memory requirements for the computing tasks. Each CCD-quarter is identified with the parent CCD number (0 \ldots 7) and a letter (k,l,m,n), as shown in the Figure \ref{fig:ccdquarterslm055} and 
\ref{fig:ccdquarterslm055_3}.

\begin{figure*}[htbp]
\begin{center}
    \includegraphics[width=0.40\textwidth]{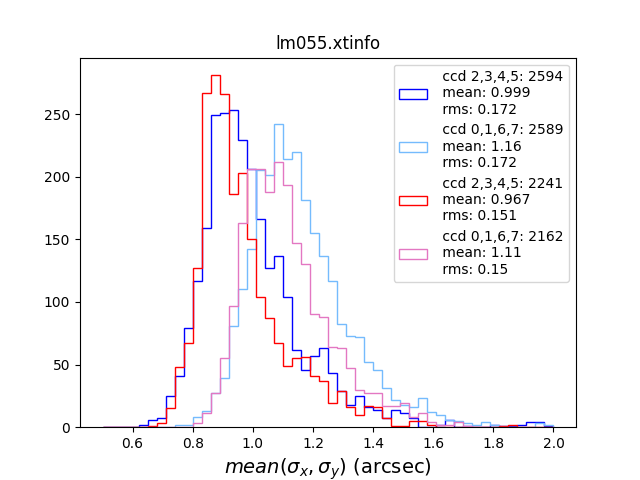} 
    \includegraphics[width=0.40\textwidth]{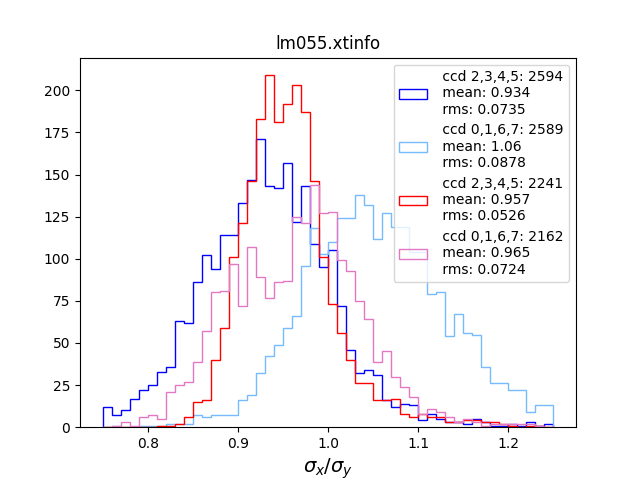}
    \caption[]{The distributions of PSF parameters, $\sigma_x$ and $\sigma_y$ defined by equation \ref{eqn:psf} for the field lm055.
    The parameters are show for the central CCDs (2,3,4,5) and peripheral CCDs (0,1,6,7) and
    for the two band passes, $\beros$ (dark and light blue) and $\reros$ (red and pink).
    }
    \label{fig:psfparameterslm055}
\end{center}
\end{figure*}

The light-curve production was carried out through the following steps

\begin{enumerate}
\item CCD bias and gain correction or ISR (Instrument Signature Removal): raw images obtained from the camera readout were corrected for a per pixel bias  and gain using master bias and gain images for each CCD. Bias images and dome flats were taken each day, before starting the sky observations, while sky flats were taken from time to time. Individual bias and gain images were processed and filtered to produce the master bias and gain used to correct the raw images. This step was performed on the EROS acquisition and processing cluster at La Silla.  
\item Reference image creation: A set of good quality exposures, typically 10 to 15 images, 
with good seeing and low level of sky background were selected to create the reference or template image through resampling and co-addition. The astrometric and geometric reference is given by the first image in the stack of co-added images. The EROS reference images are resampled with twice as many pixels ($\sqrt{2}$ along each axis) as the original image. There is one image per CCD-quarter (k,l,m,n) which covers more than a quarter of CCD, about $\sim 1200 \times 1200 \, \mathrm{pixels}$ to help managing the shift between images during co-addition. 
However, in order to avoid having the same celestial object appearing in different sub fields, each 
object catalog in this public release covers exactly a quarter of CCD,
corresponding to a $\sim 1450 \times 1450$ pixels area on the reference image, or $1024 \times 1024$ on a standard image.\footnote{The division of each ccd into quarters used for the public catalogs differs slightly than the division used for the published \erosII~papers so the designation of objects in the publications may differ from those in the public catalogs.}
It should be noted that the number of photometric measurements drops for objects near to the quarter edges, 
due to the shift in position between images, up to several hundred pixels.
\item Reference catalog (StarList) creation: Different source or star detectors  were available in the PEIDA package. The one used for the creation of the reference catalog used a filtered image obtained by the convolution of the reference image with a  Gaussian shape expected for a star or point source. The sources were then selected after applying a threshold on the filtered image, followed by steps to drop multiple detection of a single source or try to separate neighboring sources. The two starlists, obtained from the red and blue color reference images, were then cross-matched and merged, supplementing each list with sources too faint to be selected in the first color, but bright enough to be detected in the second color. We have not used directly the luminosity or flux obtained on the reference image as the flux of the object in the reference catalog, as the resampling process did not insure photon or flux conservation. The reference flux was instead obtained through the computation of the average flux computed on all the individual images used in the co-addition.
\item Each individual image quarter, called {\em current image} hereafter, was first analyzed to determine the geometrical projection of the reference catalog on it, and determine its PSF parameters. A star detector with a high threshold was used to obtain a list of rather bright sources on the image, followed by their position and flux determination. This list was then cross matched with the bright sources of the reference catalog, leading to the determination of the parameters of the geometric projection of the reference catalog on the corresponding quarter of the image. The PSF parameters on the image were then determined through PSF fitting on a sample of bright, isolated stars. PEIDA included several analytical forms for the PSF, however none was able to describe faithfully the EROS images PSF for all seeing conditions. A two dimensional Gaussian shaped PSF defined by the three parameters $\sigma_x , \sigma_y, \rho$ has been used for the standard EROS light curve production, and the finite size of pixels were taken into account in the fit. With such a Gaussian PSF, the photons from a normalized source (unit flux) centered at $(x_0,y_0)$, are spread out according to:
\begin{displaymath}
   \phi(x,y)  =  \frac{\sqrt{1 - \rho^2}}{2 \pi \sigma_x \sigma_y} \, \exp \left[ - \frac{1}{2} \left( u^2 + v^2 - 2 \rho \, u v \right) \right] 
   \end{displaymath}
   \begin{equation} \hspace*{25mm}
   u = \frac{x - x_0}{\sigma_x} \hspace{4mm} ; \hspace{4mm} v = \frac{y - y_0}{\sigma_y}.
   \label{eqn:psf}
\end{equation}
Distributions of the seeing parameters $\sigma_x$ and $\sigma_y$ for the $\approx2500$ exposures
of the field lm055 are shown in Fig. \ref{fig:psfparameterslm055}.
The effective seeing is clearly worse for the peripheral CCDs.
The mean of $\rho$ over all exposures and CCDs is near zero but its value depends on
the hour-angle and on the CCD, giving an overall r.m.s. spread of $\approx 0.15$.
\item Forced Photometry: Once the geometric projection parameters of the reference catalog on the current image has been determined, as well as the corresponding PSF parameters, they are used for the forced photometry step on the current image to compute the brightness or flux for each source and add a new point to the light curve. For each star in the reference catalog, a patch with typically 200-400 pixels around its projected position on the current image is extracted. The flux of the source is then fitted, taking into account contributions from all the neighboring source, as well as a sky background level, constant over the patch.
All the sources had their positions fixed, corresponding to the reference catalog positions, projected on the current image, as well as the PSF parameters, while fitting all the contributing sources fluxes. In addition to the fitted luminosity in each color, the fit procedure provided additional quantities, such as the flux uncertainty, and fit quality or $\chi^2 / ddf$ {\tt (chi2r, chi2b)} which are included in the data being made public. The original EROS light curve data, stored in {\tt suivi} files contains additional information, such as the fitted sky background, alternate flux determinations, aperture photometry for instance, or an independent 
estimate of the source position on the current image, which were less used in the EROS analysis and have not been included in the released data set.   
\item Relative flux alignment: although the PSF photometry corrects to some extent the observed luminosity variations due to some of the observing conditions, namely the seeing and sky background variations, the fluxes obtained from the forced photometry step on individual images are impacted by atmospheric absorption and subtle camera response variations, not corrected by the ISR. The last step of the standard photometric reconstruction in EROS corresponded to the determination of an overall absorption affecting the source luminosities. A global flux change with respect to the reference flux, modeled usually as a  linear function, was determined assuming that the major fraction of sources had a constant luminosity over time. 
The values of fluxes included in the light curves being made public {\tt (fluxr, fluxb)} and corresponding errors {\tt (efluxr, efluxb)} are the  ones obtained after correcting for a possible differential absorption with respect to the reference flux. 
\end{enumerate}

\subsection{Post-processing}
\label{sect:postprocessing}

Before the light-curve analysis was started, we removed from the light curves measurements
taken under far from normal conditions. This happened not infrequently, as the
data taking policy was to work whenever possible, leaving to the analysis
the task of rejecting these abnormal measurements. These were identified by 
extreme values of the sky background, seeing or absorption. The precise cuts 
on these quantities depended on the field but
generally  images with seeing greater than 4 arcsec were eliminated, as
well as images with absorption relative to the reference image greater than 5.
Measurements near the full moon were generally eliminated.
In addition,
images where the photometry failed for over 40\% of cataloged stars were 
eliminated, as well as images for which more than 12\% of the stars showed an
excursion from their average flux larger than three standard deviations.
Depending on the CCD, the number of rejected images varied between 
7 and 18\%, with an average of 11\%.\footnote{
The largest single cause for rejection 
was the malfunction that affected 5 CCDs of the $\reros$ passband camera 
starting in January 2002. We have thus chosen to reject all measurements 
taken in 2002 and 2003 for these 5 CCDs.}

To reduce systematic errors in the photometry,
each light curve was searched for significant
linear correlations between the measured flux and three observational 
variables, the seeing, the hour angle and the airmass. This was done 
independently in the two passbands.  
The measured fluxes were corrected linearly by requiring a vanishing
correlation coefficient between the corrected fluxes and the given 
variable. 
The largest correlation was found with the seeing, in both passbands.

\begin{figure*}[htbp]
\begin{center}
    \includegraphics[width=0.48\textwidth]{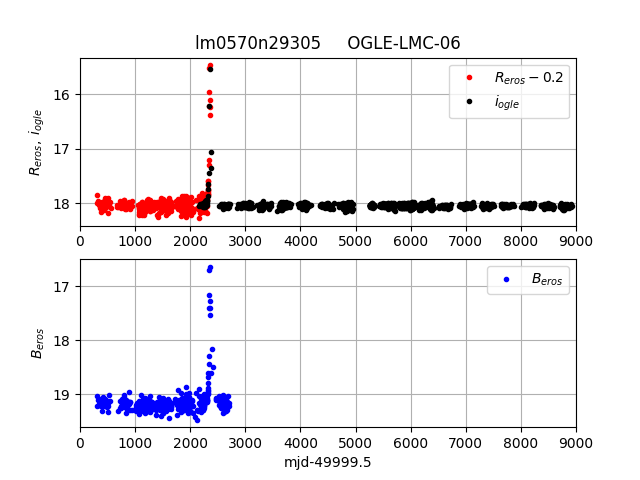}
    \includegraphics[width=0.48\textwidth]{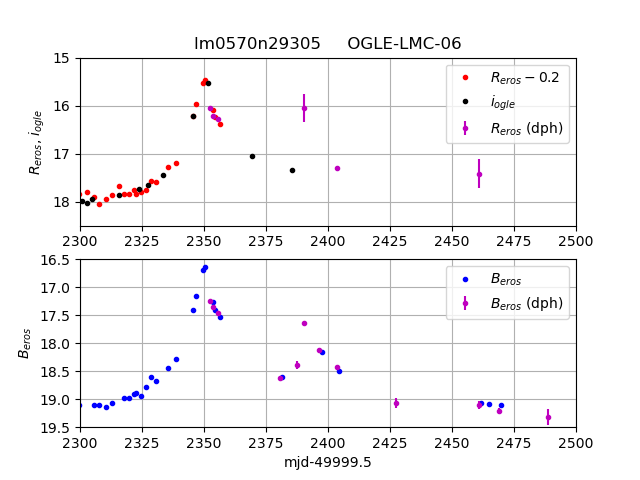}
    \caption[]{The OGLE microlensing candidate OGLE-LMC-06 as seen by \erosII \ (object lm0570n29305) and OGLE. The left figures shows EROS \citep{2007A&A...469..387T} and OGLE \citep{Ogle_2024ApJS..273....4M} magnitudes measured as a function of $(\mathrm{ehjd} = \mathrm{mjd}-49999.5)$, spanning nearly 25 years in total, 7 years for EROS. The closeup view over 200 days around the microlensing event is shown on the right. Top plots show the measurements in $\reros$ and $I_{ogle}$ bands, while $\beros$  is shown in the bottom figures. Measurements obtained through difference image photometry {\it (dph)} using the Triton package \citep{leguillou2003} have also been added on the right panels, for $\mathrm{ehjd} \gtrsim 2350$, confirming the presence of a secondary peak around $\mathrm{ehjd} \sim 2400$. To ensure readability, the {\it dph} points, in magenta, have been shifted by one day. Note that the second luminosity peak occurred near the full moon, which is the reason for the many missing points in the standard light curve.}
    \label{fig:eventlm057}
\end{center}
\end{figure*}

\begin{figure*}[htbp]
\begin{center}
    \includegraphics[width=0.48\textwidth]{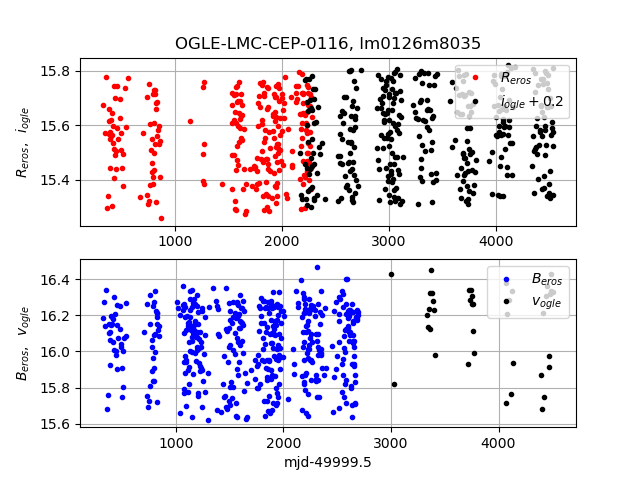}
    \includegraphics[width=0.48\textwidth]{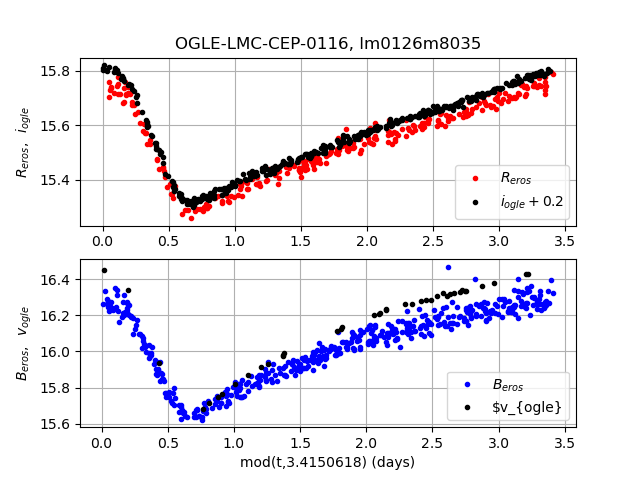}
    \caption[]{A cepheid variable star light curve , with a period of $\sim 3.4$ days, as seen by \erosII \, and OGLE. The left plots show the magnitude measurements as a function of time, in days $(\mathrm{mjd}-49999.5)$,  over more than 10 years, while the right plots show the 
    magnitude measurements folded over a single period. Top plots show the measurements in $\reros$ and $I_{ogle}$ bands, while $\beros$ and $V_{ogle}$ are shown in the bottom figures. OGLE data for this cepheid were obtained from \href{https://ogledb.astrouw.edu.pl/~ogle/OCVS/}{https://ogledb.astrouw.edu.pl/\textasciitilde ogle/OCVS/}.} \label{fig:eventogleceph}
\end{center}
\end{figure*}

\begin{figure*}[htbp]
\begin{center}
    \includegraphics[width=0.48\textwidth]{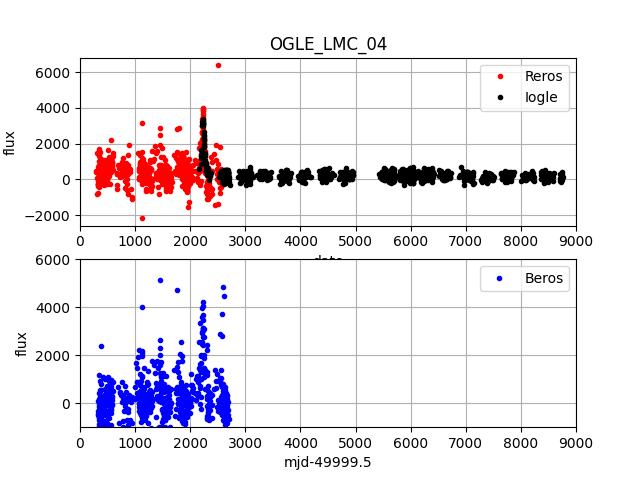}
    \includegraphics[width=0.48\textwidth]{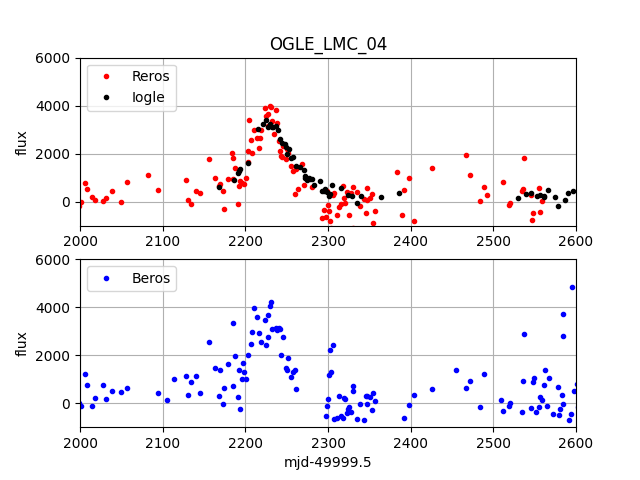}
    \caption[]{OGLE microlensing candidate OGLE-LMC-04 in OGLE \citep{Ogle_2024ApJS..273....4M} and EROS. The star is not in the \erosII~catalog so the EROS measurements have been obtained through difference image photometry using the Triton package \citep{leguillou2003} on \erosII \ images. Measurements span about 25 years, out of which $\sim 8$ years for EROS. The right figure show a closeup view over 600 days, around the microlensing event. }
    \label{fig:eventogle04}
\end{center}
\end{figure*}

Examples of light curves are shown in Figs. \ref{fig:eventlm057},  \ref{fig:eventogleceph} and \ref{fig:eventogle04}.  Figures  \ref{fig:eventlm057} and \ref{fig:eventogleceph} show a LMC microlensing event, respectively a LMC Cepheid variable star seen by both \erosII \ and OGLE. 
Figure \ref{fig:eventogle04} shows a LMC microlensing event discovered by OGLE \citep{Ogle_2024ApJS..273....4M} on a star not cataloged by \erosII.  The event was subsequently recovered through differential photometry on \erosII \ images, using the Triton software tool \citep{leguillou2003}. 

\begin{figure*}[htbp]
\begin{center}
    \includegraphics[width=0.37\textwidth]{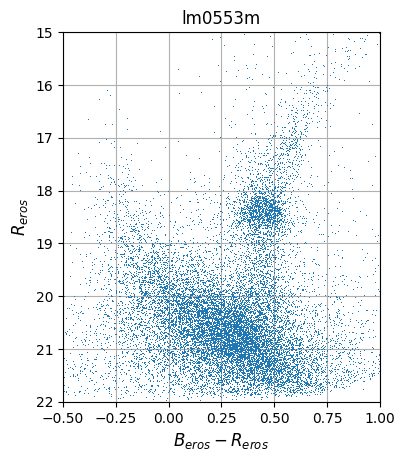} 
    \hspace{6mm} 
    \includegraphics[width=0.37\textwidth]{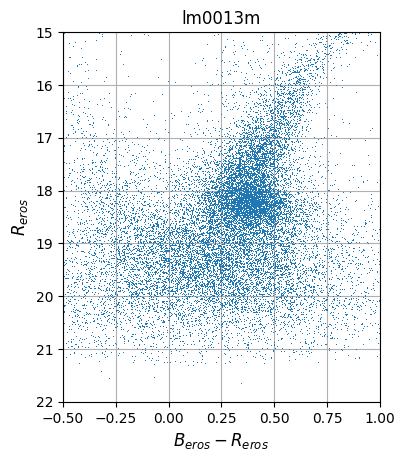} 
    \caption[]{
    The color-magnitude diagram for two CCD-quarters (CCD 3, quarter m), illustrating the observed stellar population and its variation across the LMC area. lm055 shown on the left is a typical field, while lm001 displayed on on the right is a dense field in the LMC bar. \label{fig:cmd_lm055_001}
    }
\end{center}
\end{figure*}

\begin{figure*}[htbp]
\begin{center}
    \includegraphics[width=0.40\textwidth]{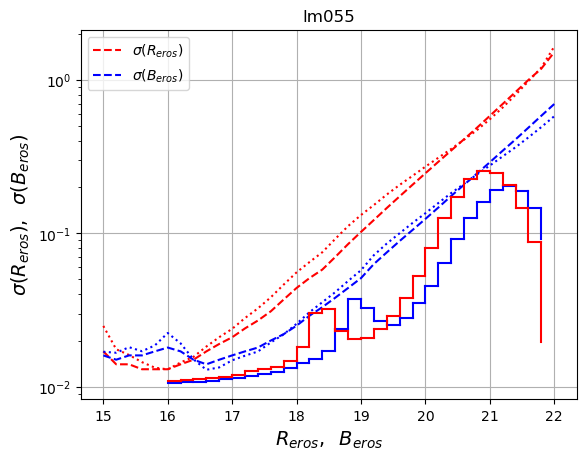} 
    \hspace{4mm} 
    \includegraphics[width=0.40\textwidth]{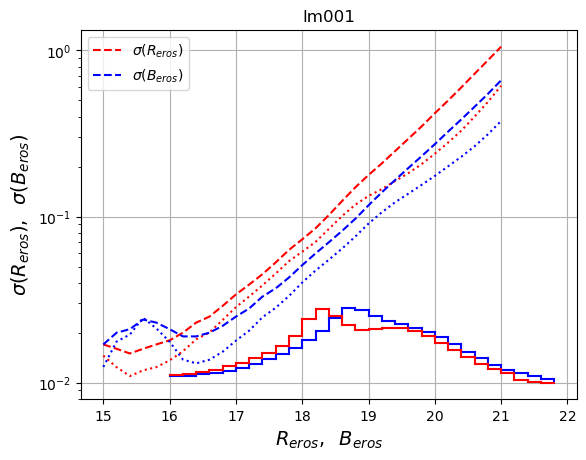}
    \caption[]{
    The magnitude distributions and photometric resolution  for the field lm055 
    (left, a field of typical density) and the field lm001  (right, a very dense field). 
    The histograms show the distribution of $\beros$ in blue, $\reros$ in red on a linear vertical scale, 
    while the lines correspond to the r.m.s. magnitude resolutions as a function of $\reros$ and $\beros$ in logarithmic scale. The  dashed blue and the dashed red lines represent photometric uncertainties for $\beros$, respectively $\reros$, as estimated by the PEIDA fitter, while the adjacent blue and red dotted lines are derived from the light-curve dispersion of the data points. 
\label{fig:emag_lm055_001}
    }
\end{center}
\end{figure*}

\begin{figure*}[htbp]
\begin{center}
    \includegraphics[width=0.42\textwidth]{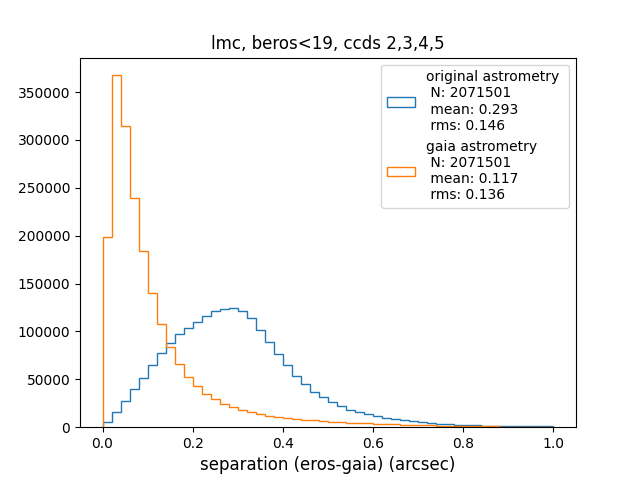}
    \includegraphics[width=0.42\textwidth]{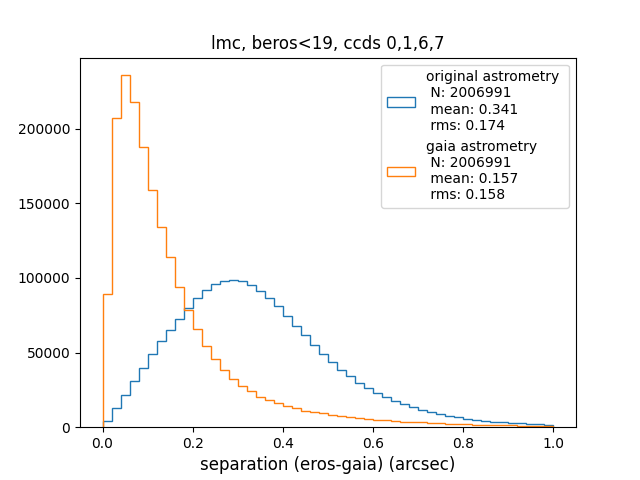}
    \caption[]{
    EROS-Gaia separations for the original eros astrometry and for the {\it Gaia} \citep{gaia_2023A&A...674A...1G} astrometry.
    }
    \label{fig:eros_gaia_separations}
\end{center}
\end{figure*}

\section{Photometric and astrometric precision}
\label{section:quality}

Two examples of LMC color-magnitude diagrams (CMD) are shown in Figure \ref{fig:cmd_lm055_001}.
They are both characterized by a clearly defined {\em bump} of clump giants. 
The main-sequence is seen in the CMD for lm055, a field of typical stellar density.
It is much less present in the dense field lm001 where the magnitude limit is lower,
due to the high density of bright stars. 

The observed stellar population magnitude distribution for the same two LMC CCD-quarters, and the associated photometric errors are shown in the Figure \ref{fig:emag_lm055_001}. 
The bump due to clump giants is clearly visible in the lm055 magnitude distribution histograms at $\reros\approx18.5$ and $\beros\approx19$.  

The photometric precision of individual measurements is nominally given by the PEIDA fitter.
The median photometric PEIDA errors as a function of magnitude are shown by the solid lines in Fig. \ref{fig:emag_lm055_001} for two fields. 
For comparison, the dotted lines show the photometric precision derived from the medians of the r.m.s. dispersion of the measurements for the entire light curve.
For lm055, the photometric precision $\sigma_\mathrm{mag}$ is better than 0.1 (10\%) 
for stars brighter than $\beros \lesssim 19.5$ in blue and  $\reros \lesssim 19$ in red. 
In very crowded fields like lm001, the same precision is reached for 0.5 magnitude brighter stars. 

The star sky coordinates of \erosII~stars were derived near the end of the survey data taking period using the astrometric resources available at the time.
Comparison with modern {\it Gaia} astrometry \citep{gaia_2016A&A...595A...1G,gaia_2023A&A...674A...1G} reveals offsets of the order 0.1-0.3 arcsec varying from field
to field.
Figure \ref{fig:eros_gaia_separations} histograms the separation of bright stars ($\beros<19$) in the eros catalog from the nearest {\it Gaia} object for central and peripheral CCDs.
The blue histogram is found using the original \erosII~astrometry, while the red histogram is
found after adjusting the \erosII~(r.a.,dec.) grid to the {\it Gaia} grid, using a quadratic fit
for each CCD-quarter.
The corrected sky coordinates using {\it Gaia} astrometry have been included in the source catalogs ({\tt .xcat} files ) being released (see paragraph \ref{sec:appdx-xcat} below).
These corrected \erosII~sky coordinates improves the mean EROS-Gaia separation from
$\approx0.3$~arcsec to $\approx0.13$~arcsec.

\section{The data set description}
\label{section:dataset}

The EROS data which is being made public through CDS is organized as a set of four types of interconnected tables. The precise definition of the quantities in the tables is given
in Appendix \ref{section:fielddesc} and examples are shown in Fig. \ref{fig:file_examples}.
The four table types are the following:
\begin{enumerate}
\item List of all object catalogs, {\tt catlist.xcatinfo}.  
For each object catalog, the list gives the 
corresponding CCD-quarter, the number of objects, the limits of the sky region, the number
of objects with better than 10 percent photometric resolution and the corresponding magnitudes,
and the magnitude limits encompassing  90 percent of the objects.
Also given are the mean offsets for the CCD-quarter between the original eros astrometry
and {\it Gaia} Data-Release 3 astrometry \citep{gaia_2023A&A...674A...1G}.
Detailed description of the fields in this general catalog can be found in section \ref{sec:appdx-xcatinfo}.
\item The Object catalogs, one for each CCD-quarter with {\tt .xcat} extension.  It gathers the information extracted from the Reference Catalog for each object or source, uniquely associated with a string identifier {\tt (erosid)}, made from the EROS field, the CCD and its quarter, concatenated with an integer sequence number. For each entry, the table contains  the source position on the sky and on the reference image, its luminosity as obtained by the photometric reconstruction from images stacked to compute the reference image. Each table entry (a source) is enriched with quantities computed from the corresponding light curve: the average and median magnitude and associated fluctuations. This data set is organized as one file for each CCD quarter and for each field, with {\tt .xcat} extension and the file content is described in section 
\ref{sec:appdx-xcatinfo}.
\item Light curves correspond to the sequence of photometric measurements and the associated date/time for each object or source in the object catalog. Each light curve is stored as a single file with {\tt .xtime} extension and the source {\tt erosid} as the file name.
The content of the light curve files is described in section \ref{sec:appdx-xtime}.
\item The image catalogs, with one file per EROS field with {\tt .xtinfo} extension, contains the basic information for each of the exposures taken on that field.  
There is one entry for each image corresponding to one CCD in a camera, the image itself being stored as a FITS file. In addition to the file name and exposure date/time encoded as the Modified Julian Day (MJD), this table contains synthetic information about the image quality provided by the photometric reconstruction, such as the Gaussian PSF parameters. Note that despite the processing by CCD-quarter, yielding image parameters for each quarter, the xtinfo files provide only per CCD quality parameters. Although the images are not directly stored at the CDS, they are accessible through links from the EROS database at the CC-IN2P3. 
The content of the image catalog files is described in section  \ref{sec:appdx-xtinfo}.
\end{enumerate}

\section{Conclusions}
We have presented in this paper the database of the \erosII~program available through the CDS portal.
It is hoped that the light curves and images will be useful for understanding transient objects observed in present and future programs.
\section{Data availability}
The data set described in this paper is available in electronic form at the CDS  via \href{https://cdsarc.cds.unistra.fr/viz-bin/cat/II/390}{https://cdsarc.cds.unistra.fr/viz-bin/cat/II/390}. Detailed information on the electronically available tables is provided in the appendix \ref{section:fielddesc}.

\newpage
\begin{acknowledgements}
In addition to the authors, the production and analysis of the \erosII~data relied on
the work of our deceased colleagues:  Pierre Bareyre, Florian Bauer, Sergio Char, Jean Guibert, Jacques Haissinski, Eric Maurice, Alain Milsztajn, Luciano Moscoso, and C\'ecile Renault.
We are appreciative of their contribution and dedicate this publication to their memory.
We thank Jean-Baptiste Marquette for his work on the astrometry
and Jean-Paul Le F\`evre for help with the production of the catalogs.
This work has made use of data from the European Space Agency (ESA) mission
{\it Gaia} (\url{https://www.cosmos.esa.int/gaia}), processed by the {\it Gaia}
Data Processing and Analysis Consortium (DPAC,
\url{https://www.cosmos.esa.int/web/gaia/dpac/consortium}). Funding for the DPAC
has been provided by national institutions, in particular the institutions
participating in the {\it Gaia} Multilateral Agreement. 

This research made use of hips2fits,\footnote{https://alasky.cds.unistra.fr/hips-image-services/hips2fits} a service provided by CDS.

\end{acknowledgements}

\bibliographystyle{aa}
\bibliography{aa57321-25}
\nocite{*}

\begin{appendix}
\section{Detailed description of \erosII~table columns}
\label{section:fielddesc}
\begin{figure*}[htbp]
\begin{center}
    \includegraphics[width=\textwidth]{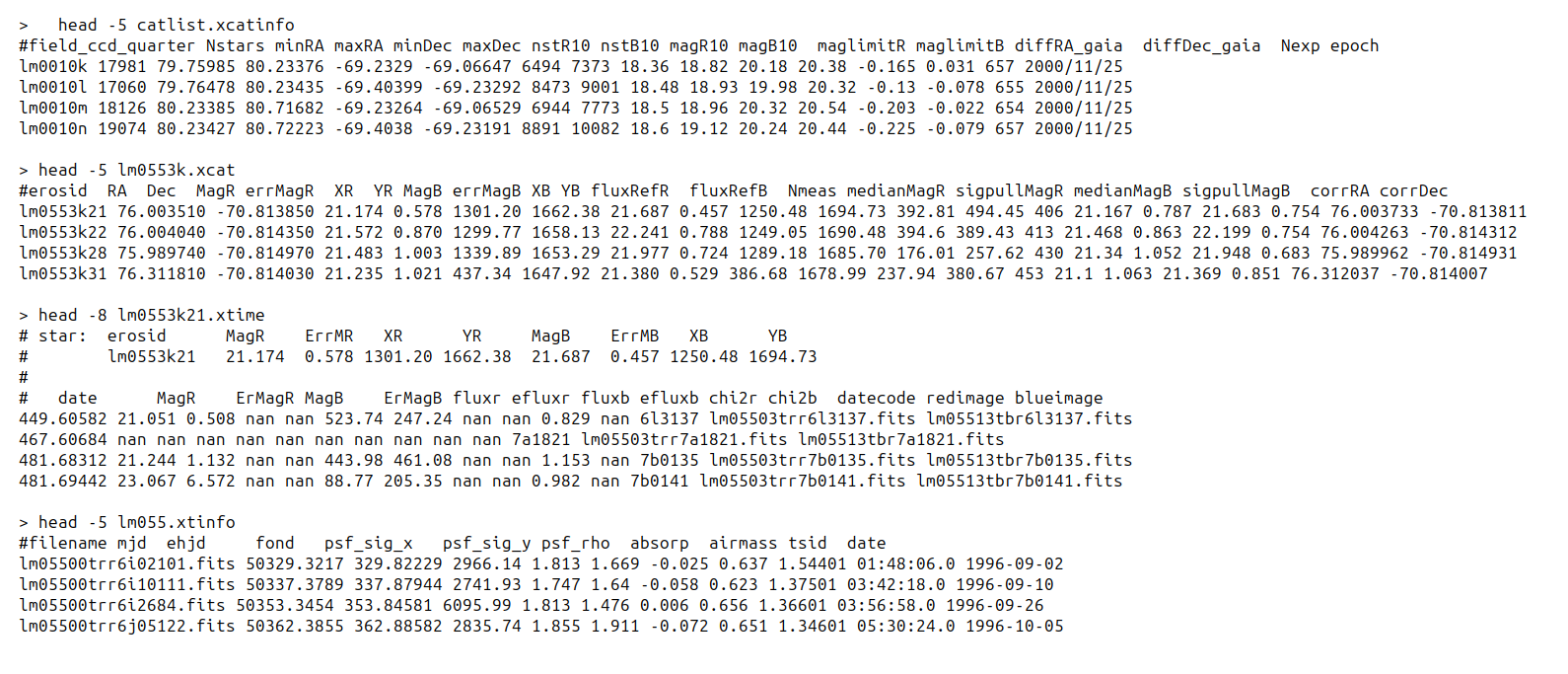}
    \caption{The top lines of examples of the tables described in Appendix A.
    From top to bottom: the list of {\tt xcat} files, {\tt catlist.xcatinfo};
    the object catalog ({\tt xcat}) file for the CCD-quarter {\tt lm0553k};
    the light-curve ({\tt xtime}) file for the object {lm0553k21}; and
    the image list ({\tt tinfo}) file for the field {lm055}.
    }
    \label{fig:file_examples}
\end{center}
\end{figure*}

In this appendix we describe the elements of the publicly-available files, examples
of which are shown in Fig. \ref{fig:file_examples}. All the files described here are
plain text files, with one header line starting with the hash character, with the name
of the fields in each line of the file. The values on each line, as well as the field 
of column names in the first line are separated by spaces. 

These files have been transferred to CDS and ingested as VizieR tables, and accessible through the EROS data set URL: \href{https://cdsarc.cds.unistra.fr/viz-bin/cat/II/390}{https://cdsarc.cds.unistra.fr/viz-bin/cat/II/390}. It can be queried interactively through the VizieR web portal\footnote{CDS VizieR web portal: \href{https://vizier.cds.unistra.fr}{https://vizier.cds.unistra.fr} }  
and subset of the EROS data can be selected and downloaded using python packages such as PyVO\footnote{PyVO :  \href{https://pyvo.readthedocs.io/en/latest/}{https://pyvo.readthedocs.io/en/latest/} } or VizieR Queries\footnote{VizieR Queries (astroquery.vizier) python package: \\  \href{https://astroquery.readthedocs.io/en/latest/vizier/vizier.html}{https://astroquery.readthedocs.io/en/latest/vizier/vizier.html} }. 

The catalog and light curve files described here are archived at CC-IN2P3 and can be accessed through web links (URL) which will be displayed on the EROS collection pages at CDS. 
\subsection{List of \erosII~catalogs (xcatinfo)} 
\label{sec:appdx-xcatinfo}
This single file named {\tt catlist.xcatinfo} presents a top level view of the whole data set, One with one line containing the following information for each xcat file.
\begin{itemize}
\item \apxbf{field\_ccd\_quarter} (string) : The field identification, followed by the CCD number (0-7) and quarter (k,l,m,n). For example {\tt sm0023m} refers to the SMC field 002, CCD 3 and quarter m.  
\item \apxbf{Nstars} (integer) : Number of stars in the corresponding object catalog (xcat file).
\item \apxbf{minRA, maxRA, minDec, maxDec} (float) Limit in right-ascension and declination covered by the corresponding object catalog.
\item \apxbf{nstR10, nstB10} (integer) : Number of stars with better than 10 percent photometric precision in $\reros$ and $\beros$ bands. 
\item \apxbf{magR10, magB10} (float) : Upper magnitude limits for stars with photometric precision better 10 percent  in $\reros$ and $\beros$.
\item \apxbf{maglimitR, maglimitB} (float) : Upper magnitudes limits encompassing 90 percent of stars in $\reros$ and $\beros$.
\item \apxbf{diffRA\_gaia, diffDec\_gaia} (float) : Mean offset of original \erosII~astrometry compared to {\it Gaia} astrometry \citep{gaia_2023A&A...674A...1G} in right-ascension and declination.
\item \apxbf{Nexp} (integer) Number of exposures or measurements for the corresponding xcat file.
\item \apxbf{epoch} (string) The observation date of the first image used to create the red band reference image, formatted as {\tt yyyy/mm/dd}. This defines the epoch of the celestial source positions on the image plane that have been used to determine the astrometric projection.
\end{itemize}
Values of most of the above quantities, aggregated over all fields for each target, 
is given in table \ref{tab:listoftargets}

\subsection{\erosII~Object catalog (xcat)} 
\label{sec:appdx-xcat}
This part of the EROS data set gathers object catalog files, with one file 
for each CCD-quarter with one line per object. There are a total of 
2783 files for the EROS LMC fields (lm), 316 for SMC fields (sm), 2532 files 
for the Galactic Bulge (cg) fields and 913 for spiral arm fields (gs/bs/gn/tm). 
The Object catalog gathers the information for each source or star for which a light curve is provided, organized as up to 32 files for each EROS-2 field, one for each CCD quarter. The file names are made of the field identifier, followed by the CCD number and the quarter identifier letter, with the file extension {\tt .xcat}. 
The file {\tt lm0554m.xcat} for example contains the list of sources for the LMC field 055, CCD \# 4, quarter m. The two letters {\tt lm, sm, cg, bs, gs, gn, tm} are the prefixes used for the LMC, SMC, Galactic Bulge, $\beta$Sct, $\gamma$Sct, $\gamma$Nor and $\theta$Mus respectively.
  
\begin{itemize}
\item \apxbf{erosid} (string) : unique EROS object identifier, composed of the field name, CCD number and quarter and a integer, a sequence number identifying the source in the corresponding sub-field. For example, {\tt lm0554m4824} is the erosid for the source with the sequence number 4824, in quarter m of CCD 4, in the field lm055. 
\item \apxbf{RA, Dec} (float) :  The source position on the sky, the right ascension and the declination $(\alpha, \delta)$ in decimal degrees and J2000 epoch according to the original \erosII~astrometry. 
\item \apxbf{MagR, MagB} (float):  Source apparent magnitude in the EROS red ($\reros$) and blue ($\beros$) passbands. These values are computed from measurements decorrelated from observing conditions  (Sect. \ref{sect:postprocessing}) along 
the light curve, so they cannot be obtained directly from {\tt fluxRefR, fluxRefB} values 
described below. 
\item \apxbf{errMagR, errMagB} (float):  uncertainty of the source magnitudes for the two passbands $(\reros,\beros)$ 
\item \apxbf{XR, YR, XB, YB} (float):  The source position along the first (X) and second (Y) CCD axis on the red (R) and blue (B) reference images, in pixel units. 
\item \apxbf{fluxRefR, fluxRefB} (float): source luminosity or flux, in ADU units, on the red $\reros$ and blue $\beros$ EROS passbands in the reference catalog. These fluxes correspond to the average of fluxes determined by the forced photometry on the set of images stacked to obtain the reference image. 
\item \apxbf{Nmeas} (integer) :  number of measurements on the light curve (no unit) 
\item \apxbf{medianMagR, medianMagB} (float):  median magnitudes computed from the light curve in the EROS red and blue passbands. 
\item \apxbf{sigpullMagR, sigpullMagB} (float): standard deviation of magnitudes with respect to the median in the two passbands
\item \apxbf{corrRA, corrDec} (float) :  Corrected source position on the sky, using {\it Gaia} astrometry\citep{gaia_2023A&A...674A...1G}. The right ascension and the declination $(\alpha, \delta)$ are expressed in decimal degrees and J2000 epoch. 
\end{itemize}

\subsection{\erosII~Light curves (xtime)}
\label{sec:appdx-xtime}
Each light curve, composed of the sequence of the photometric measurements and the corresponding date and time for a given celestial source is stored in a file having the EROS source identifier (\apxbf{erosid}) as its name, with the extension {\tt .xtime}. 
There is one xtime file for each star in the xcat.
Each xtime begins with a four-line header followed by one line per exposure.
The header contains the \apxbf{erosid}, \apxbf{(MagR,MagB)}, \apxbf{errMagR, errMagB} and \apxbf{(XR,YR,XB,YB} from the corresponding xcat line.

The following information is provided for each point along the light curve.
{\em Note that a {\tt  nan } (not a number) is written when the data is missing.}
\begin{itemize}
\item \apxbf{date} (float):  EROS heliocentric Julian date ($\approx mjd-49999.5$), identical to the 
\apxbf{ehjd} field of the image catalog {\tt (.xtinfo)}. 
\item \apxbf{MagR, MagB} (float):  Estimated magnitudes in the two EROS passbands $(\reros,\beros)$ for this point along the light curve.  
\item \apxbf{ErMagR, ErMagB} (float):  Estimated magnitude uncertainties in the two passbands. 
\item \apxbf{fluxr, fluxb} (float) : source luminosity in the two passbands, in ADU units, as computed by the photometric pipeline on the two associated images, corrected by the relative photometric alignment procedure. 
\item \apxbf{efluxr, efluxb} (float): luminosity or estimated flux uncertainties, in ADU's, as computed by the photometric reconstruction. 
\item \apxbf{chi2r, chi2b} (float):  Reduced chi-square $(\chi^2/n_{dof})$ of the PEIDA photometric PSF fit in the two passbands - can be used to drop unreliable flux estimates.
\item \apxbf{datecode} (string): The encoded exposure date (year,month,day), followed by an exposure sequence number used in the image names. The date, encoded with 4 characters, correspond to the day before the night observations (same date in UTC or local time). The first character (0..9a..z) represent the year, starting with 0 for 1990. 6 corresponds thus to 1996, a to 2000, c to 2002. The second character (a..l) represents the month, and two characters, the day as two digits.   
\item \apxbf{redimage} (string): The name of the image in the $\reros$ band. 
\item \apxbf{blueimage} (string): The name of the image in the $\beros$ band. 
\end{itemize}

Note that for each point along the light curve, the source luminosities \apxbf{fluxr, fluxb} as estimated by the photometric pipeline have been slightly corrected, using a linear function to cancel the observed average biases correlated with image seeing, the hour angle (target observation angle with respect to the meridian) and the airmass, as described in \citep{tisserand:tel-00008267}, chapter 6, paragraph 6.3. The seeing is the parameter having the largest effect on the photometric biases. 
The corrected luminosities have then been converted into magnitudes, using zero point magnitudes derived from the absolute photometric calibration.

\subsection{\erosII~Image catalog (xtinfo)}
\label{sec:appdx-xtinfo}

This catalog provides basic characteristics of EROS images used to obtain the light curves. It is organized as one file per field, with the corresponding field identifier as the file name with the {\tt .xtinfo} extension. The file {\tt lm055.xtinfo} contains for example the list of all images for LMC field lm055 with one entry per CCD. The table has the following columns described below. \\[1mm]
\begin{itemize}
\item \apxbf{filename} (string) : image file name which is obtained from a concatenation of the EROS  field, the camera and CCD numbers, the image type (reduced image in this case), the encoded exposure date (year,month,day) (the day before the night observations)  and an exposure sequence number for that observation night. Note that these are the offset subtracted and gain corrected images. Refer to the datecode field in {\tt .xtime} files in paragraph \ref{sec:appdx-xtime} for the description of date and sequence number encoding in the image names. 
\item \apxbf{mjd} (float): exposure date and time represented as the modified Julian date MJD\footnote{Modified Julian Day: MJD = JD - 2400000.5 } (unit=days).

\item \apxbf{ehjd} (float): EROS heliocentric Julian date, in days unit, identical to the time stored in light curve files. It is obtained from the \apxbf{mjd}, correcting for the light propagation time if the telescope was moved to the center of the solar system, with an offset=49999.5, as if the origin was shifted to early October 1995.
\item \apxbf{fond} (float) : sky background corresponding to typical pixel brightness level, for image area outside sources, in ADU unit. 
\item \apxbf{psf\_sig\_x, psf\_sig\_y, psf\_rho}  (float):  Gaussian PSF parameters, in pixel units (1 pixel $\simeq 0.6^"$) 
\item \apxbf{absorp} (float) : the absorption relative to reference image. 
This parameter is obtained from the distribution of the ratio 
$\frac{\mathrm{fluxCal}}{\mathrm{fluxUncal}}$
of typical stars flux after relative photometric alignment ($\mathrm{fluxCal}$) divided by the one before   ($\mathrm{fluxUnCal}$).
\item \apxbf{airmass} (float): 
$\sim 1/\cos(za)$ with $za$ the zenith angle.
\item \apxbf{tsid} (string) : Local sidereal
time, formatted as {\tt hh-mm-ss.s}
\item \apxbf{date} (string) : the exposure date, formatted as {\tt year-month-day}. Note that this is the string representation of the date encoded in the image name, which corresponds 
to the day before the night observations.   
\end{itemize}

\section{\erosII~images}
\label{sec-eros2-images}
The reference co-added images, projected to CDS-Aladin grid are available at CDS and can be viewed using the Aladin viewer.  
The reference images, as well as individual images are archived at the CC-IN2P3 an can be accessed through links displayed by the CDS VizieR 
when browsing through the image catalog.
The reference images correspond to resampled, stacked image, with a pixel size $(\sim 0.4^")$ smaller that the EROS camera CCD pixels, with each image covering more than a quarter of a CCD, with $1800 \times 1800$ pixels, corresponding to a $\sim 1270 \times 1270$ pixels area on the CCD.  Standard images, referred also to as {\em current} images in this article, covers the full CCD with $2048 \times 2048$ pixels.

\section{\erosIccd~Data}
\label{sec-eros1-data}
The data from \erosIccd~survey of Magellanic clouds, carried out with the T40 \citep{1994ExA.....4..265A} setup from 1991 to 1995 is also being made public.
The 40 cm reflecting telescope was equipped with a mosaic of 16 buttable CCD's, each with $400 \times 579$ pixels, covering a total area of $1^\circ \times 0.4^\circ$ on sky. A single field in the LMC bar (Fig. \ref{fig:erosfieldsmap}) was observed with high cadence, during three observing seasons, 1991-92, 1992-93 and 1993-94. A single field in SMC was observed in the 1994-95 season. Images were successively taken through two filters, defining the EROS-1 red (R) and blue (B) bands. The interest of this data set is mainly related to the high sampling rate of the light curves, less than half an hour in each band.   
The analysis for each observing season was carried out independently. Light curves were obtained with an early version of the photometric reconstruction pipeline described in section \ref{section:3dataprocessing}. 

\subsection{\erosIccd~Images}
\label{sec-eros1-images}
For each observing season, reference co-added images as well as individual images for each CCD are being made available, one co-added image for each CCD and each of the two filters (R,B), and around 1000 images for each CCD and each filter. Note that there are missing CCD's in each season. The reference images were made independently for each CCD, 
with an oversampling factor 2.5. A narrow band around the image was cut out, to ensure a uniform exposure, despite small shifts between individual images, resulting in images of approximately $\sim 900 \times 1400$ pixels, covering slightly less than a single CCD, with the exact reference image size depending on the observing season.

An image catalog, similar to the one described in section \ref{sec:appdx-xtinfo}, organized as one table for each season provides characteristics of the standard images. A separate file provides the list of the reference image for the observing season. 

\begin{itemize}
\item \apxbf{filename} (string) : image file name which is obtained from a concatenation of several parts, starting with {\tt xe} for LMC and {\tt se} for SMC, followed by three characters, CCD number {\tt (00 01 \ldots 15)}, and the filter {\tt (b,r)}. The last 6 characters of the name correspond to the 
encoded exposure date (the day before the night observations) as (year,month,day, 4 characters) and an exposure sequence number in hexadecimal for that observation night. 
The file {\tt xe04r2b211e.fits} correspond to an LMC image, CCD 04, with the red filter, taken on 21 February 1992, and a sequence number {\tt 1e}. 
\item \apxbf{NumCCD} (integer): CCD number , 0 \ldots 15
\item \apxbf{Filter} (character): filter (B / R) 
\item \apxbf{RA, Dec} (float) : image center position 
\item \apxbf{DateTimeTU} (string) : observation date and time, formatted as {\tt yyyy-mm-ddThh:mm:ss.s } for example {\tt 1991-12-19T10:07:11.0}
\item \apxbf{TStart} (integer) : observation time in seconds, since January 1$^\mathrm{st}$, 1990, at 00h.
\item \apxbf{Fond} (float) : sky background corresponding to typical pixel brightness level, for image area outside sources, in ADU unit. 
\item \apxbf{psf\_sig\_x, psf\_sig\_y, psf\_rho}  (float):  Gaussian PSF parameters, in pixel units (1 pixel $\simeq 0.6^"$) 
\item \apxbf{absorp} (float) : the absorption relative to reference image, computed from the distribution of the ratio of typical star fluxes, after and before relative photometric alignment. 
\item \apxbf{airmass} (float): $\sim 1/\cos(za)$ with $za$ the zenith angle.
\end{itemize}

\subsection{\erosIccd~star catalog and light curves}
A catalog for a subset of sources for which the light curves are available is provided for each of the four observing seasons. As for EROS-2, photometric calibration was performed using the Magellanic Clouds Photometric Survey  \citep{zaritsky}. \erosIccd~magnitudes in the two filters can be converted   to standard magnitudes through the relations:
\begin{eqnarray*}
\berosccd  \simeq  0.7 \Vj + 0.3 \Bj \hspace{10mm} 
\rerosccd  \simeq  0.3 \Vj + 0.7 \Ic \\
\berosccd, \rerosccd = -2.5 \log10( fluxB,R ) + Z_p^{B,R}  
\end{eqnarray*}
with zero point magnitudes $Z_p^B \simeq 25.25$, $Z_p^R \simeq 24.75$.

The list of objects for which light curve are available can be found the object catalog file, with one file for each CCD and each observing season. The file {\tt LMC9293\_14.xcat} for example gathers the list of stars for the 1992-93 observing season 
and CCD 14. 
The object catalog contains the following fields: 
\begin{itemize}
\item \apxbf{erosid} (string) : unique EROS object identifier, composed of one character identifying the target, {\tt x} for LMC, {\tt s} for SMC, 
CCD number {\tt 00, 01 \ldots 15} and quarter and a sequence number identifying the source in the corresponding CCD. 
\item \apxbf{RA, Dec} (float) :  The source position on the sky, the right ascension and the declination $(\alpha, \delta)$ in decimal degrees and J2000 epoch. 
\item \apxbf{MagR, MagB} (float):  Source apparent magnitude in the EROS-1 red  and blue passbands. 
\item \apxbf{XR, YR, XB, YB} (float):  The source position along the first (X) and second (Y) CCD axis on the red (R) and blue (B) reference images, in pixel units. Note that the images publicly accessible have had their first axis inverted (see above, paragraph \ref{sec-eros1-images}) 
\item \apxbf{fluxRefR, fluxRefB} (float): source luminosity or flux, in ADU units, on the red  and blue  EROS-1 passbands in the reference catalog. 
\item \apxbf{NmeasR, NmeasB} (integer) :  number of measurements in the red and blue light curves (no unit) 
\item \apxbf{medianFluxR, medianFluxB} (float) : median flux in the two passbands, along the light curve. 
\item \apxbf{sigpullFluxR, sigpullFluxB} (float) : dispersion of the flux values in the two passbands, with respect to the median along the light curve. 
\end{itemize}

For each source in the object catalog, there is one light curve file, gathering the photometric measurements as a function of time in the two passbands. Each light curve contains the following fields:
\begin{itemize}
\item \apxbf{TStart} (integer) : observation time in days, since January 1$^\mathrm{st}$, 1990, at 00h. 
\item \apxbf{Filter} (character): filter (B / R) 
\item \apxbf{flux} (float) : source luminosity in ADU units, as computed by the photometric pipeline on the two associated images, corrected by the relative photometric alignment procedure. 
\item \apxbf{errflux} (float): luminosity or estimated flux uncertainties, in ADU's, as provided by the photometric reconstruction. 
\end{itemize}

\section{\erosIplates~light curves}
\label{sec-eros1-schmidtplate-data}

The \erosIplates~program used observations during three annual periods (October-March),
the first plate taken October 16,1990 and the last March 27, 1993.
A total of 290 usable photographic
$29\times29$ cm$^2$ plates were obtained at the E.S.O
Schmidt telescope (1 meter free aperture, f/3).
Half of the plates (098-04 emulsion) were taken with a RG630 red filter
and half (IIaO emulsion) with a GG385 blue filter.
Exposure times were 1 hour in each color, and apart from the
very crowded LMC bar region,
our star detection efficiency abruptly drops at a limiting
magnitude of around 20.5 in red and 21.5 in blue.
The data taking period is limited by the maximum
excursion of the telescope around the meridian position
($\pm2.5$ hours).
The time sampling of the plates 
makes the experiment sensitive to
microlensing event durations ranging from a few days to a few months.

The usable field shown in Fig \ref{fig:erosfieldsmap} covered with a plate is $5.2^{\circ}\times5.2^{\circ}$,
centered
on position $(\alpha = 5h20',\delta = -68^{\circ}30')$ (eq. 2000).
The transmission coefficient, $T$, of each plate was digitized to 0.8
giga-pixels of 10 microns size (0.675 arcsec), with 12 bits dynamic range, using the
``MAMA"\footnote{MAMA was developed and operated by INSU/CNRS.}
microdensitometer
(Machine Automatique \`a Mesurer pour l'Astronomie) at the
Observatoire de Paris (\citep{Berger}). The digitization took 6 hours per
plate, yielding $28 \times 28 = 784$ images, covering each $\sim 1 cm^2$ 
with $1024 \times 1024$ pixels.
The Schmidt plates are being digitized again at the Observatoire de Paris and the resulting digital images will be made available through CDS.

Photometry was performed using the quantity $\Phi_T=[(1-T)/T]^{0.6}$ which varies
approximately linearly with the flux $\Phi$ collected on the
photographic plate in the magnitude range $[17,21]$ (blue)
$[16,20]$ (red).
We therefore use $\Phi_T$ in our photometric fitting and star finding
procedures. The thus obtained star magnitudes were originally
corrected using an empirical formula 
derived from comparison of selected fields with CCD
measurements taken with the E.S.O-Danish 1.54m telescope.
Subsequent comparison with the Zaritsky catalog \citep{zaritsky}
has confirmed the calibration to a precision of 0.2mag outside the LMC bar.
Stars in the densest regions of the LMC bar may have listed magnitudes
incorrect by as much as 1~mag.

For analysis purposes, the LMC field was divided into 784 sub-fields covering 1~cm$^2$
of the Schmidt plate.  These sub-fields are referred to as {\em pav\'es}.
The list of objects for which light curve are available can be found the object catalog file, with one file for each {\em pav\'e}. The file {\tt pave660.xcat} for example gathers the list of stars for the pav\'e 660.

The object catalog contains the following fields: 
\begin{itemize}
\item \apxbf{erosid} (string) : object identifier within the  given pav\'e. 
\item \apxbf{RA, Dec} (float) :  The source position on the sky, the right ascension and the declination $(\alpha, \delta)$ in decimal degrees and J2000 epoch. 
\item \apxbf{NmeasR, NmeasB} (integer) :  number of measurements in the red and blue light curves (no unit) 
\item \apxbf{MagR, MagB} (float):  Mean source apparent magnitude in the EROS-1 (Schmidt) red  and blue passbands. 
\item \apxbf{errMagR, erMagB} (float):  Mean source apparent magnitude  error in the EROS-1 (Schmidt) red  and blue passbands. 
\item \apxbf{sigpullMagR, sigpullMagB} (float) : dispersion of the flux values in the two passbands, with respect to the median along the light curve. 
\end{itemize}

For each source in the object catalog, there is one light curve file, gathering the photometric measurements as a function of time in the two passbands. Each light curve contains the following fields:
\begin{itemize}
\item \apxbf{TStart} (integer) : observation time in days, since January 1$^\mathrm{st}$, 1990, at 00h. 
\item \apxbf{Filter} (character): filter (B / R) 
\item \apxbf{Mag} (float) : source luminosity in ADU units, as computed by the photometric pipeline on the two associated images, corrected by the relative photometric alignment procedure. 
\item \apxbf{errMag} (float): luminosity or estimated flux uncertainties, in ADU's, as provided by the photometric reconstruction. 
\item \apxbf{Fond} (float): Sky background under the star. 
\item \apxbf{Poll} (float): "pollution" from neighboring stars. 
\end{itemize}
\end{appendix}

\end{document}